\documentclass[11pt,a4paper]{article}
\pdfoutput=1
\usepackage{jheppub}

\usepackage{ifpdf}
\usepackage{afterpage}
\usepackage{amssymb}
\usepackage{amsmath}
\usepackage{graphicx}
\usepackage{longtable}
\usepackage{verbatim}
\usepackage{amsfonts}
\usepackage{bbold}
\usepackage[table,usenames,dvipsnames]{xcolor}
\usepackage{titlesec}
\usepackage{enumerate}
\usepackage{gensymb}
\usepackage{url}
\usepackage{fontawesome}

\usepackage{tikz}
\usetikzlibrary{trees}
\usetikzlibrary{decorations.pathmorphing}
\usetikzlibrary{decorations.markings}
\tikzset{
    photon/.style={decorate, decoration={snake}, draw=black},
    wino/.style={draw=redwine},
    electron/.style={draw=black, postaction={decorate},
        decoration={markings,mark=at position .55 with {\arrow[draw=black]{>}}}},
    scalar/.style={draw=black, dashed,postaction={decorate},
        decoration={markings,mark=at position .55 with {\arrow[draw=black]{>}}}},
    gluon/.style={decorate, draw=black,
        decoration={coil,amplitude=4pt, segment length=5pt}}
}

\newcommand{\Sec}[1]{Sec.~\ref{#1}}

\newcommand{\App}[1]{Appendix~\ref{#1}}
\newcommand{\Tab}[1]{Table~\ref{#1}}
\newcommand{\Fig}[1]{Fig.~\ref{#1}}
\newcommand{\Refe}[1]{Ref.~\cite{#1}}

\newcommand{\Eq}[1]{Eq.~(\ref{#1})}
\newcommand{\Eqs}[2]{Eqs.~(\ref{#1}) and (\ref{#2})}

\newcommand{\ie}{\textit{i.e.}\ }

\newcommand{\beq}{\begin{equation}}
\newcommand{\eeq}{\end{equation}}
\newcommand{\bea}{\begin{eqnarray}}
\newcommand{\eea}{\end{eqnarray}}
\newcommand{\nn}{\nonumber}

\def\OMIT#1{{}}
\newcommand{\lsim}{\mathrel{\rlap{\lower4pt\hbox{\hskip1pt$\sim$}}
    \raise1pt\hbox{$<$}}}         
\newcommand{\gsim}{\mathrel{\rlap{\lower4pt\hbox{\hskip1pt$\sim$}}
    \raise1pt\hbox{$>$}}}         

\newcommand{\bl}{\left}
\newcommand{\br}{\right}

\newcommand{\dd}{\mathrm{d}}
\newcommand{\DP}{\ensuremath{\mathrm{D P}}}

\newcommand{\ee}{\mathrm{e}}
\newcommand{\so}{\mathrm{s}}
\newcommand{\pk}{\mathrm{pk}}
\newcommand{\tr}{\mathrm{tran}}
\newcommand{\ad}{\mathrm{ad}}
\newcommand{\fr}{\mathrm{free}}
\newcommand{\ec}{\mathrm{echo}}

\newcommand{\atm}{\mathrm{atm}}
\newcommand{\cmb}{\mathrm{cmb}}

\newcommand{\GeV}{\mathrm{GeV}}

\newcommand{\Hz}{\mathrm{Hz}}

\newcommand{\GHz}{\mathrm{GHz}}

\newcommand{\cm}{\mathrm{cm}}

\newcommand{\kpc}{\mathrm{kpc}}

\newcommand{\Jy}{\mathrm{Jy}}
\newcommand{\lumcgs}{\mathrm{erg \cdot s^{-1} \cdot \Hz^{-1}}}

\newcommand{\ignore}[1]{}

\newcommand{\githubmaster}{\href{https://github.com/ManuelBuenAbad/snr\_ghosts}{\faGithub}}

\vspace*{-30mm}

\title{\boldmath Axion Echoes from the Supernova Graveyard}
\author[a,b,e]{Manuel A. Buen-Abad,}
\author[b,c]{JiJi Fan,}
\author[d]{Chen Sun}
\affiliation[a]{Maryland Center for Fundamental Physics, University of Maryland, College Park, MD, 20742, USA}
\affiliation[b]{Department of Physics, Brown University, Providence, RI, 02912, USA}
\affiliation[c]{Brown Theoretical Physics Center, Brown University,
Providence, RI, 02912, U.S.A.}
\affiliation[d]{School of Physics and Astronomy, Tel-Aviv University, Tel-Aviv 69978, Israel}
\affiliation[e]{Dual CP Institute of High Energy Physics, C.P. 28045, Colima, M\'{e}xico}
\emailAdd{buenabad@umd.edu}
\emailAdd{jiji\_fan@brown.edu}
\emailAdd{chensun@mail.tau.ac.il}

\vspace*{1cm}

\abstract{Stimulated decays of axion dark matter, triggered by a source in the sky, could produce a photon flux along the continuation of the line of sight, pointing backward to the source. The strength of this so-called axion ``echo" signal depends on the entire history of the source and could still be strong from sources that are dim today but had a large flux density in the past, such as supernova remnants (SNRs). This echo signal turns out to be most observable in the radio band. We study the sensitivity of radio telescopes such as the Square Kilometer Array (SKA) to echo signals generated by SNRs that have already been observed, and show that SKA could reach axion-photon couplings of order $g_{a\gamma\gamma} \sim \mathcal{O}(10^{-11}) \,\mathrm{GeV}^{-1}$ for axion masses $m_a \lesssim 10^{-5}\;\mathrm{eV}$. In addition, we show projections of the detection reach for signals coming from old SNRs and from newly born supernovae that could be detected in the future. Intriguingly, an observable echo signal could come from old ``ghost" SNRs which were very bright in the past but are now so dim that they haven't been observed. 
\githubmaster
}

\begin{document}
\maketitle

\abstract{Abstract}


\section{Introduction}

An axionlike particle, $a$, is a periodic pseudoscalar field, $a \cong a + 2 \pi f_a$ with $f_a$ the decay constant. The discrete shift symmetry acting on the field is its defining feature. It protects the small mass of the axion, $m_a$, and imposes strong constraints on the forms of its couplings. Axions arise ubiquitously from top-down theory such as string models~\cite{Svrcek:2006yi}, and low-energy phenomenological models~\cite{Peccei:1977hh, Peccei:1977ur, Weinberg:1977ma, Wilczek:1977pj, Kim:1979if, Shifman:1979if, Zhitnitsky:1980tq, Dine:1981rt}. They serve as one of the most motivated scenarios of feebly coupled light particles beyond the standard model. In particular, an important observational target is the axion's coupling to photons in the standard model, which is usually parametrized as
\begin{equation}
-\frac{g_{a\gamma\gamma}}{4} a F_{\mu\nu} \tilde{F}^{\mu\nu} = g_{a\gamma\gamma}a \mathbf{E} \cdot \mathbf{B}, 
\label{eq:aphoton}
\end{equation} 
where the coupling $g_{a\gamma\gamma}$ has energy dimension $-1$, inversely proportional to $f_a$; $F_{\mu\nu}$ is the field strength of electromagnetic $U(1)_{\rm em}$; $\tilde{F}^{\mu\nu} = \frac{1}{2} \epsilon^{\mu\nu\alpha\beta} F_{\alpha\beta}$ is the dual field strength, and $\mathbf{E}, \mathbf{B}$ are the corresponding electric and magnetic fields.

Another attraction of an axion is that it could serve as a cold dark matter candidate~\cite{Preskill:1982cy, Dine:1982ah, Abbott:1982af}. The simplest canonical mechanism is misalignment: an axion is initially frozen at some random place in its field space due to the Hubble friction. When the Hubble expansion rate drops around its mass, the axion starts to oscillate around the minimum of its potential. The coherent oscillations redshift as cold matter and store energy in the axion field. 

\begin{figure}[h]
\centering
\includegraphics[width=.8\textwidth]{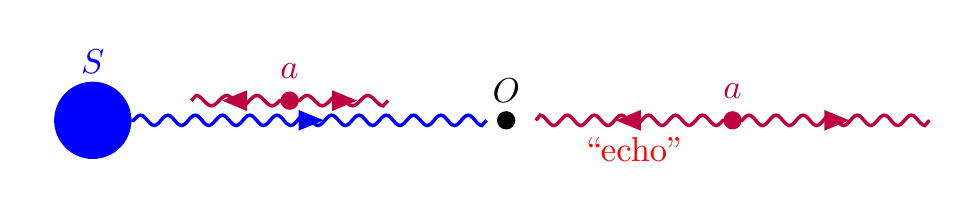}
\caption{Schematic illustration of the axion echo signal. $S$ and $O$ stand for the source and the observer respectively while $a$ indicates axion dark matter. Wavy lines represent photons, from either the source (blue) or stimulated decays of axion dark matter (purple). The photon from stimulated decays, along the continuation of the line of sight to the direction opposite to the source, is the echo signal.}\label{fig:echo}
\end{figure}

Given its theoretical importance, there has been a rapidly growing interest in searching for axions and their couplings over the years, cf.~\cite{Graham:2015ouw,CAST:2007jps,CAST:2017uph,ADMX:2009iij,ADMX:2018gho,ADMX:2019uok,ADMX:2021nhd,ADMX:2018ogs,Bartram:2021ysp,Crisosto:2019fcj,PhysRevLett.59.839,PhysRevD.42.1297,Lee:2020cfj,Jeong:2020cwz,CAPP:2020utb,HAYSTAC:2020kwv,Alesini:2019ajt,Alesini:2020vny,Foster:2020pgt,Darling:2020uyo,Battye:2021yue,Bernal:2020lkd,Bloch:2021vnn} to name a few. In this article, we will study a novel technique to probe the coupling of axion dark matter to photons, which has been underexplored, that is, searching for an {\it ``echo"}\footnote{This has also been called ``axion \textit{Gegenschein}'' \cite{Ghosh:2020hgd}.} signal from stimulated axion dark matter decays.\footnote{The axion dark matter's decay lifetime is still much longer than the age of the universe so it could still be the dominant component of dark matter.} The basic idea is depicted in Fig.~\ref{fig:echo}: photons from a source traverse the axion dark matter halo. If the photon energy matches half of the axion mass, it induces stimulated decays of axion dark matter into two photons. Due to momentum conservation, these two photons travel in opposite directions with energies equal to half of the axion mass, $m_a/2$. An observer could receive two fluxes of photons from opposite directions. One flux is along the line of sight from the source to the observer, which is a superposition of the continuum emission of the source (blue wavy line) and the line emission from axion decays with an angular frequency set by $m_a/2$ (purple wavy line). Since the line emission from the stimulated decays is a much fainter signal, it is challenging to isolate it from the bright source continuum background. The other flux is the so-called ``echo" signal also from stimulated axion decays, which is along the continuation of the line of sight to the direction {\it opposite to the source}. This one could potentially be a clean signal if there is no bright source in the opposite direction.

\begin{figure}[th]
  \centering
  \includegraphics[width=1.0\textwidth]{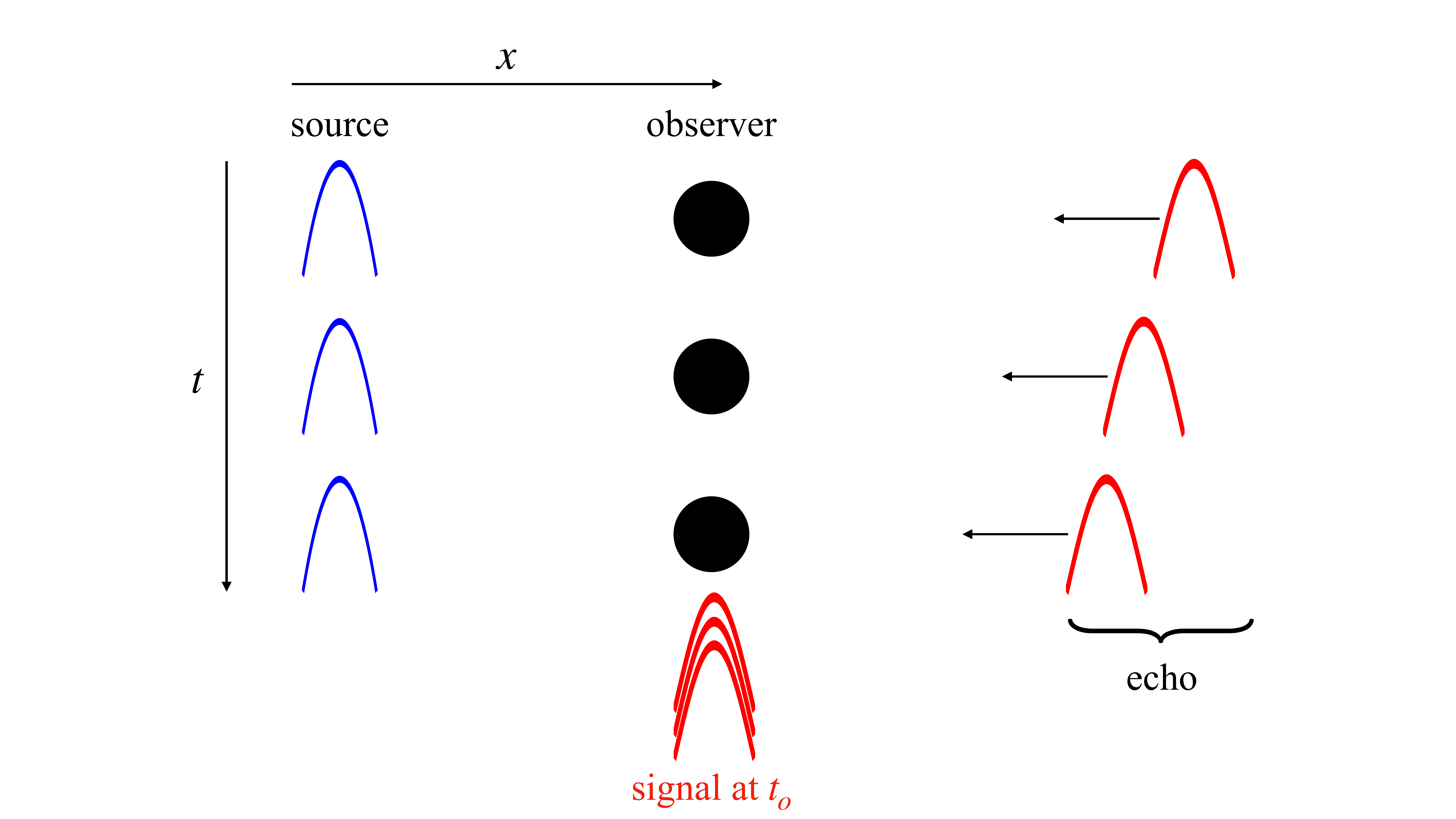}
  \caption{Illustration that the echo signal observed depends on the entire history of a time-dependent source. The source emits a series of pulses (blue) over time. They travel past the observer and trigger stimulated decays of axion dark matter, resulting in echo photons (red). While the photons comprising the signal arrive at the observer at the same time $t_o$, they are associated with pulses emitted from the source at different times.}
  \label{fig:illustration}
\end{figure}

The echo technique was first applied and developed in Refs.~\cite{Arza:2019nta, Arza:2021nec}, with the source being a powerful radio beam shooting from the Earth. Given the experimental challenges, it will be helpful to identify natural sources in the universe. So far the only astrophysical source that has been examined in the literature to trigger an echo signal is Cygnus A (Cyg~A), an extragalactic active galactic nucleus~\cite{Ghosh:2020hgd}. In this work, we point out for the first time that the echo signal relies on the historical luminosity of the photon source, while the signal of photons from stimulated decays of axion dark matter that travel along with the source photon as studied in Ref.~\cite{Caputo:2018vmy,Ayad:2020fzc,Chan:2021gjl} require the brightest radio sources {\it today}. This opens up the possibility to probe axion DM with {\it time-varying} radio sources that may be dim today but were once very bright, such as supernova remnants (SNRs).

The analyses of SNRs will also be drastically different from Cyg A, and any other relatively stable sources over the timescale for light to travel across the dark matter halo. An SNR, as a time-dependent source, typically starts with a huge photon flux from the explosion of a supernova and undergoes several different phases before eventually merging with the interstellar medium. In this case, the echo signal could be induced by the flux from the SNR at time much earlier than that at which the observation is made. For a more intuitive understanding, one could consider an SNR emitting a series of pulses over time, as shown in the vertical direction of the cartoon in Fig.~\ref{fig:illustration}. The pulse emitted at earlier times could pass the observer and travel further before triggering the decay of axion dark matter and generating an echo photon, while the later-emitted pulse travels a shorter distance past the observer before triggering the axion stimulated decays, as demonstrated by the horizontal direction in Fig.~\ref{fig:illustration}. These echo photons could travel back and arrive at the observer at the same time $t_o$. In other words, the observed signal is a collection of the echo photons generated by the pulses from the source at different times. More concretely, as we will show in a more rigorous way, the signal is an integration {\it over the history of the SNR}. It turns out that even a dim source today could still lead to an observable echo signal, since it was brighter earlier.

Throughout the paper, we work with natural units. 
Our paper is organized as follows. In Sec.~\ref{sec:axion-dark-matter}, we derive our master formula to compute the axion echo signal from a time-dependent source, and comment on different effects of relative motions between SNR, the observer, and the dark matter halo. In Sec.~\ref{sec:snr}, we review the basics and time evolution of SNRs that we will use in our analysis. In Sec.~\ref{sec:detection}, we discuss the radio telescope, Square Kilometer Array, used for detecting the echo signal and the calculation of the signal to noise ratio. We present our results in Sec.~\ref{sec:results} and conclude in Sec.~\ref{sec:conclusion}. Finally, in \App{app} we check the uncertainties in the sensitivity of Square Kilometer Array to axion echoes trigged by the most promising SNR, G39.7-2.0. We have made publicly available the Python code we developed to aid our calculations; it can be found at \href{https://github.com/ManuelBuenAbad/snr\_ghosts}{\tt github.com/ManuelBuenAbad/snr\_ghosts}.

\section{Echo signal}
\label{sec:axion-dark-matter}

In this section, we will first derive the key formula to compute the flux density (also referred to as spectral irradiance in the literature), $S_{\nu}$, of the echo signal from a time-dependent source such as SNRs. We will then comment on the effects of relative motions between the source, the observer, and dark matter halo.

\subsection{Flux density of echo signal}
\label{sec:fluxdensity}

Since we will consider SNRs in our galaxy, we ignore the expansion of the universe. The particle number density $n$ is related to its phase space density $f(\mathbf{p})$ as $dn = \DP f$, with $\DP \equiv g \dd^3p/(2\pi)^3$ and $g$ the degrees of freedom of the particle. In our scenario, the evolution of the photon phase space density $f_\gamma$ is governed by the Boltzmann equation:
\begin{align}
\frac{\dd}{\dd t} f_\gamma
  & =
    C[f_\gamma] \ ,
    \label{eq:boltzmann}
\end{align}
where $C[f_\gamma]$ is the collision term for photons. Taking into account both the stimulated decays of axion dark matter and the inverse process, $a(\mathbf{p_a}) \rightarrow  \gamma(\mathbf{p}_1) +  \gamma(\mathbf{p}_2)$ and $  \gamma(\mathbf{p}_1) +  \gamma(\mathbf{p}_2) \rightarrow a(\mathbf{p}_a) $, the collision term is given by:
\begin{align}
    C[f_1] = \frac{1}{2E_1} \int\!\frac{\DP_a}{2E_a}\int\!\frac{\DP_2}{2E_2} \ \vert \overline{\mathcal{M}} \vert^2 \ \Big (f_a(1+f_1+f_2) - f_1 f_2 \Big ) \ (2\pi)^4 \delta^4(p_a - p_1 - p_2) \ ,
    \label{eq:collision}
\end{align}
where $E$'s ($p$'s) are the energies (momenta) of the particles involved (subscripts $a$, $1$ and $2$ denote axion, photon 1 and 2 respectively); $f$'s are the phase space densities; and $\vert \overline{\mathcal{M}} \vert^2$ is the averaged matrix element squared of the processes. 

It can be shown, for the axion and photon occupation numbers in our study (which are functions of the dark matter density, the luminosity of the source, and the axion mass and photon energy), that the backward process $\propto f_1 f_2$ is negligible compared to that of the forward process $\propto f_a(1+ f_1 + f_2)$. Furthermore, in a leading-order approximation, we take the axion dark matter to be perfectly cold and thus we can ignore its motion, \ie $f_a = n_a (2\pi)^3 \delta^3(\mathbf{p}_a)$. For photons coming from directional sources, the phase space density peaks in a given direction, $f_1 = f_\gamma(|\mathbf{p}_1|) h(\hat{\mathbf{p}}_1)$, with the direction function $h(\hat{\mathbf{p}}_1)$ in general anisotropic and $\hat{\mathbf{p}}_1$ being the direction of the original photon propagation. For photons from point sources such as SNRs, $h(\hat{\mathbf{p}}_1) \approx \delta^2(\Omega_1 \hat{\mathbf{p}}_1 + \Omega_* \hat{\mathbf{n}}_*)$, where $\Omega_1$ is the corresponding phase space solid angle of $\hat{\mathbf{p}}_1$, and $\Omega_*$ is the solid angle of the light source as seen from the Earth, in the direction $\hat{\mathbf{n}}_*$. Said differently, for point sources the phase space solid angle $\Omega_1$ is that of the light source $\Omega_*$, and the direction of motion of the photons $\hat{\mathbf{p}}_1$ is (obviously) \textit{antiparallel} to the vector $\hat{\mathbf{n}}_*$ going from the observer to the source.

Finally, due to momentum conservation, integrating Eq.~\eqref{eq:boltzmann} and Eq.~\eqref{eq:collision} over $\DP_a$ and $\DP_2$ gives:
\begin{align}
\frac{\dd}{\dd t} f_\gamma
  & =
    \frac{\pi^2 \Gamma_a}{E_a^3} \rho_a
    \Big ( 1 + f_\gamma h(\hat {\mathbf{p}}_1) + f_\gamma h(-\hat {\mathbf{p}}_1 ) \Big )
    \delta(E_\gamma - E_a) \ ,
    \label{eq:fgamma}
\end{align}
where the axion (spontaneous) decay width is given by $\Gamma_a \equiv g_{a\gamma\gamma}^2m_a^3/(64\pi)$, $\rho_a$ is the dark matter energy density $\rho_a =n_a m_a$, and $E_a = m_a/2$. The three terms in the parentheses correspond to the isotropic spontaneous decay, the stimulated decay photon along the original photon direction (as mentioned above, antiparallel to $\hat{\mathbf{n}}_*$), and the echo photon going in the \textit{opposite} direction (\ie, \textit{parallel} to $\hat{\mathbf{n}}_*$, pointing toward the source), respectively. In what follows, we focus on the latter term, the echo signal.

For pointlike sources, as opposed to extended ones, the relevant observable is the flux density $S_{\nu}$, which is related to the phase space density as $S_\nu = \frac{E^3}{2\pi^2} \int \dd \Omega \; f(E, \Omega)$, with $\nu = E/2\pi$ the photon frequency. It can further be shown that the flux of the echo signal is much smaller than that of the source, \ie we stay in the perturbative response regime. In other words, the echo stimulated \textit{by an echo} (as opposed to by the original source itself) is negligible.\footnote{Cf. Ref.~\cite{Levkov:2020txo} for a study in the nonperturbative regime.} We can therefore integrate Eq.~\eqref{eq:fgamma} over the solid angle and then over time, in order to obtain the flux density of the echo signal $S_{\nu_a,\ee}$ observable today, and relate it to the source's flux density $S_{\nu_a,\so}(t)$, which is a function of time:
\begin{align}
\label{eq:delta_snu_echo}
    S_{\nu_a, \ee} = \frac{\pi^2 \Gamma_a}{E_a^3} \delta(E_\gamma - E_a) \int\limits_{0}^{t_{\rm age}/2} \! \dd x \ \rho_a(x, -\hat{\mathbf{n}}_*) \ S_{\nu_a, \so}(t_{\rm age} - 2x) \ .
\end{align}
Here $\nu_a = E_a / (2\pi) = m_a / (4\pi)$. Note the crucial retarded time argument in the source's flux density $S_{\nu_a,\so}(t)$. This reflects the fact that the portion of the echo emitted by axion dark matter at a distance $x$ along the line of sight (and being observed today, when the source has an age $t_{\rm age}$) was produced by light that first passed the observer's location a time $2x$ ago. Note also that the axion dark matter density is evaluated along the line of sight in the direction opposite to the source, hence the name of \textit{``echo''}. Throughout the rest of this paper, we will assume that the dark matter is distributed according to the Navarro-Frenk-White profile \cite{Navarro:1995iw}, with a scale radius of $20~\kpc$ \cite{Schaller:2015mua,Calore:2015oya} and a local density of $0.4~\GeV \, \cm^{-3}$ in the solar neighborhood \cite{Sivertsson:2017rkp,Buch:2018qdr}.

We can immediately see from \Eq{eq:delta_snu_echo} that, under the assumption of perfectly cold dark matter, the echo signal looks like a line at photon energy equal to $E_a = m_a/2$. Let us now relax that assumption. Indeed, the dark matter has a velocity dispersion $\sigma_v \approx 5\times 10^{-4} c$ ($c$ being the speed of light) \cite{Freese:2012xd}. As a result, the echo signal is not exactly a delta function, but a distribution with a finite width of order $\sim \sigma_v E_a$. Finally, it can be explicitly shown that for dark matter velocity distributions following the standard halo model, the echo signal is approximately distributed as a Gaussian with standard deviation $\sigma_v E_a$.

For narrow signals, the observationally relevant quantity is the flux density averaged over the bandwidth of interest $\Delta \nu$. This can be obtained from \Eq{eq:delta_snu_echo} simply by integrating the finite distribution, and dividing by $\Delta \nu$. The result is
\begin{align}
\label{eq:snu_echo}
    \overline{S}_{\nu_a, \ee} = f_\Delta \frac{\pi \Gamma_a}{2E_a^3 \Delta \nu} \int\limits_{0}^{t_{\rm age}/2} \! \dd x \ \rho_a(x, -\hat{\mathbf{n}}_*) \ S_{\nu_a, \so}(t_{\rm age} - 2x) \ \mathrm{e}^{-\tau(\nu, x, -\hat{\mathbf{n}}_*)} \ .
\end{align}
Here $f_\Delta$ is the fraction of the total (integrated) signal flux falling within the bandwidth $\Delta \nu$. Anticipating the results from \Sec{sec:detection}, we point out here that there is an optimal $\Delta \nu$ (and therefore $f_\Delta$) that will maximize the ratio of the signal to the noise, with the latter scaling as $\propto \sqrt{\Delta \nu}$. Therefore, approximating the signal distribution as a Gaussian function, we find that the optimal numbers are \cite{Ghosh:2020hgd}:
\beq
\Delta \nu  \approx 2.8 \, \nu_a \sigma_v \, , \quad f_\Delta \approx 0.84 \ .
\label{eq:bandwidth}
\eeq

In \Eq{eq:snu_echo} we have included the optical depth $\tau$. The optical depth due to free-free absorption can be computed as follows \cite{1974ApJ...194..715G}:
\begin{align}
  \tau(\nu)
  & \approx
    9.5 \times 10^{-7}
    \left ( \frac{\rm EM}{\mathrm{cm}^{-6}\, \mathrm{pc}} \right )
    \left ( \frac{T_e}{5000\;\mathrm{K}} \right )^{-1.38} 
    \left ( \frac{\nu}{\mathrm{GHz}} \right )^{-2.08} \ ,
\end{align}
where ${\rm EM} = \int n_e^2 \, d\ell$ is the emission measure of electrons along the photons' propagation, and $T_e$ is the temperature of electrons. There have been many attempts to model the electron number density in the galaxy, e.g. Ref.~\cite{Schnitzeler:2012jq}. When both the echo signal and the source photon come from a direction away from the galactic center, one could estimate the electron number density with a single scale height $h$ using a simple exponential function $n_e = n_{e,0} \mathrm{e}^{-|z|/h}$. A fit from \cite{Schnitzeler:2012jq} with pulsar measurements (model M1 therein) gives an electron density in the galactic midplane as $n_{e,0} \approx 0.015 \;\mathrm{cm}^{-3}$. This translates to an EM:
\begin{align}
      {\rm EM}
  & \approx
    0.23\;\mathrm{cm^{-6}\;pc}\,
    \left (\frac{n_{e,0}}{0.015\;\mathrm{cm}^{-3}} \right )^2
    \left ( \frac{\ell}{\mathrm{kpc}} \right ) \ ,
\end{align}
which makes the optical depth negligible for the entire range of frequency we investigate. However, when the echo signal comes from the vicinity of the galactic center, this expression greatly underestimates the EM. From Ref.~\cite{1989ApJ...342..769P}, the emission measure is estimated to be ${\rm EM}\sim 10^{5}\; \mathrm{cm}^{-6}\mathrm{pc}$, based on the radio flux from Sgr A$^*$. Using this, we can find the optical depth at the galactic center to be $\sim \mathcal{O}(1)$ at $\nu \approx 300\;\mathrm{MHz}$. Note that the emission measure in \cite{1989ApJ...342..769P} is based on the radio observation of Sgr A$^*$ and the 7~arcmin halo surrounding it. 
When echo signals come straight from the galactic center, the background is very strong leading to poor sensitivity. In what follows, we do not discuss signals coming directly from the galactic center, therefore we neglect the optical depth.

\subsection{Motions of source, observer, and dark matter}
\label{sec:moving-sources}
In the derivation above, we implicitly assume that the source, the observer on Earth, and the dark matter halo are at rest. However, in reality, they all move with respect to each other, and their relative motions could cause a reduction of the echo signal, which we will estimate in this section. For simplicity, we will study one motion at a time. The general motions of the system could be more complicated but the overall effect is a combination of the three effects discussed below. 

First, we consider the coherent motion of the source. An echo photon points back toward the position the source had at the time when it emitted the photon that triggered the stimulated decays. Since the signal is a stack of the echo photons, the motion of the source causes an {\it aberration} of the echo image. During the time relevant for generating the echo signal, the source travels a distance $d_s$. If $d_s$ is much below the distance $D$ between the source and the observer, the aberration angle is estimated to be
\begin{align}
\theta_{\rm ab} \approx \frac{d_s}{D} \approx  10 \,{\rm arcmin} \, \left(\frac{v_s}{10^{-3}}\right)\left( \frac{t_{\rm age}}{10^4 \, {\rm years}} \right ) \left ( \frac{1\;\mathrm{kpc}}{D} \right) \, ,
\label{eq:ab}
\end{align}
where $v_s$ is the speed of the source, for which we picked a characteristic value as a benchmark, considering SNRs in our galaxy. For a SNR with size $\sim \theta_{\rm ab}$, the image will be blurred and enlarged by an order one factor.

On the other hand, the motion of the Earth away from the source could lead to a {\it reduction of the flux}. The change in the total echo photon flux received could be estimated as 
\begin{align}
1-\left(\frac{D}{d_o+ D}\right)^2 \approx \frac{2 d_o}{D} = 6 \times 10^{-3} \, \left(\frac{v_o}{10^{-3}}\right)\left( \frac{t_{\rm age}}{10^4 \, {\rm years}} \right ) \left ( \frac{1\;\mathrm{kpc}}{D} \right) \, ,
\end{align}
where $d_o$ is the distance the observer travels and $v_o$ is the observer's velocity. One can see that this is a negligible effect. 

\begin{figure}[h]
\centering
\includegraphics[width=1.0\textwidth]{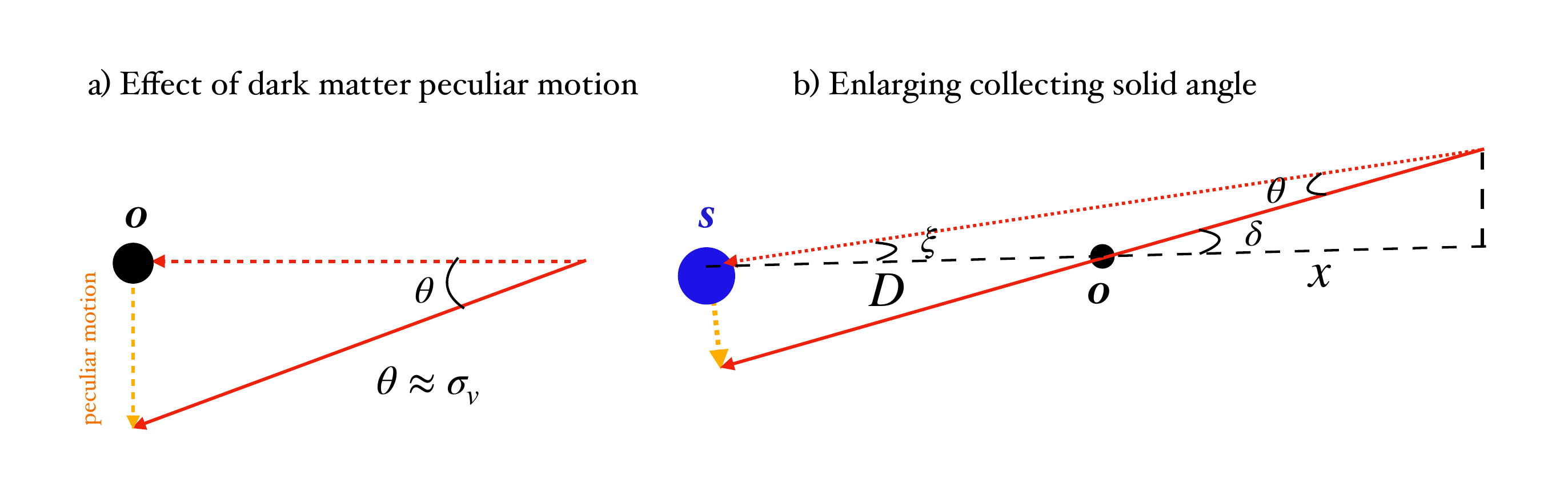}
\caption{\textit{Left:} peculiar motion (dashed orange line) could cause the echo photon to depart from its original route (dashed red line) and make it escape from being collected by the telescope at O. For illustration, we show the peculiar motion in the direction transverse to the line of sight. The final route of the echo photon is depicted by the red solid arrow. \textit{Right:} enlarging the collecting angle to $\delta$ to collect echo photons from neighboring points in order to compensate for the loss due to peculiar motions. The blue dot represents the source; the observer is shown as a black dot. More details can be found in the main text.}\label{fig:peculiar}
\end{figure}

Finally, we consider the peculiar motion of the dark matter. The peculiar motion could cause the echo photon to move out of the telescope's receiver dish, as shown in panel (a) of Fig.~\ref{fig:peculiar}. More specifically, due to the peculiar motion of dark matter particles, an echo photon could depart from its original route and move over a distance of order $\sigma_v t_{\rm age} \sim 10$ light years, for a SNR of $10^4$ years old, in the direction transverse to the line of sight. This is significantly larger than a telescope dish size and the photon could not be received. To overcome this difficulty, we need to consider a wider collecting solid angle, as illustrated in panel (b) of Fig.~\ref{fig:peculiar}. Let us consider an echo photon that could not hit the detector without peculiar motion (the original route is denoted by the dashed red line). With peculiar motion (represented by the orange dashed line; for simplicity, we assume that the peculiar motion is in the transverse direction), the echo photon will hit the detector along the red solid route. Thus increasing the collecting angle could help collect echo photons nearby to overcome the loss due to the peculiar motion. A straightforward computation based on the geometry in panel (b) tells us that as long as $\sigma_v \ll D/x$, the new collecting angle we need is given by:
\beq
2\delta \approx 2 \sigma_v \frac{x+D}{D} \approx 2 \sigma_v \frac{t_{\rm age}/2 + D}{D} \ ,
\label{eq:extendedangle}
\eeq
where, in the second step, we take the maximum distance, $x$, which an echo photon being collected today could travel, to be $t_{\rm age}/2$. Note that the aberration angle $\theta_{\rm ab}$ in Eq.~\eqref{eq:ab} and $\delta$ are of similar size since the rotational speed of a source not far away from the solar system and the dark matter peculiar velocity are of the same order.

\subsection{Shadow of the Earth}
\label{sec:shadow-observer}

While the relative motions of dark matter, source, and the observer can cause the reduction or blurring of the signal, as discussed above, they can also help overcome the effects of the shadow of the Earth. When both the source and the observer are at rest in the dark matter rest frame, the shadow of the observer, the Earth in our case, will form. The shadow could be detrimental to the signal in two ways: (a) stimulated decays of axions located within the Earth's shadow are prevented; (b) the echo photons from axion decays outside the shadow will not pass the Earth. However, neither of the two effects poses a problem due to the motion of the Earth in the dark matter rest frame.

The shadow at any given moment starts from the Earth, extending to the direction opposite to the source. Given that the angular diameter size of the sources are much larger than that of the Earth, the length of the shadow can be estimated as
\begin{align}
  \ell_{\rm sd}
  & \approx
    \frac{2R_{\oplus}}{\theta_{\so}} \ ,
\end{align}
where $R_{\oplus}$ is the radius of the Earth and $\theta_\so$ is the angular diameter size of the source. Let us take Cassiopeia A (G111.7-2.1 from Green's SNR catalog \cite{Green:2005yt,Green:2014cea,Green:2015isa,Green:2019mta}) as an example. Its associated shadow size is $\ell_{\rm sd} \approx 0.06 \;\mathrm{au}$. However, this estimate is obtained by using the current size of the SNR. In the past, the size was smaller and therefore the shadow it cast was larger. Nevertheless, even taking the SNR expansion into account, the length of the shadow barely reached $\sim 1\;\mathrm{au}$ when the SNR was around 10 years old. This leads to a negligible reduction of the integral in Eq.~(\ref{eq:snu_echo}), which has $t_{\rm age} \sim \mathcal{O}(10^{4}) \;\mathrm{year} \sim \mathcal{O}$(a few) kpc for most cases of interest.

Now let us look at the seemingly more fatal challenge from effect (b). If everything is still in the dark matter rest frame, stimulated photons trace back to the source along the exact route of the original source photons. An obvious condition for echo photons to be generated is that the source photons should not be blocked by the Earth. This leads to the simple observation that no echo photons will be captured by a telescope on the Earth regardless of either the size of the shadow or the finite size of the source if the Earth does not move! Fortunately, the Earth, together with the entire solar system, moves with respect to the dark matter halo. It only takes a negligible amount of time for the Earth to move out of its own shadow, $t\sim R_{\oplus}/v_{\odot} \sim 21 \;\mathrm{s}$, as the solar system revolves around the galactic center with a circular speed $v_\odot \sim 10^{-3}$. Thus effect (b) of the shadow is also not a problem for observing the echo signal.

\section{Supernova remnants}
\label{sec:snr}
In this section, we will discuss the properties of SNRs, as well as their time evolution, both of which will be used in our analysis. We do not intend to provide a thorough review on this vast subject and will instead refer interested readers to Refs.~\cite{2013tra..book.....W, Dubner:2015aqa,Green:2005yt,Green:2014cea,Green:2015isa,Green:2019mta}. 

\begin{figure}[th]
  \centering
  \includegraphics[width=.5\textwidth]{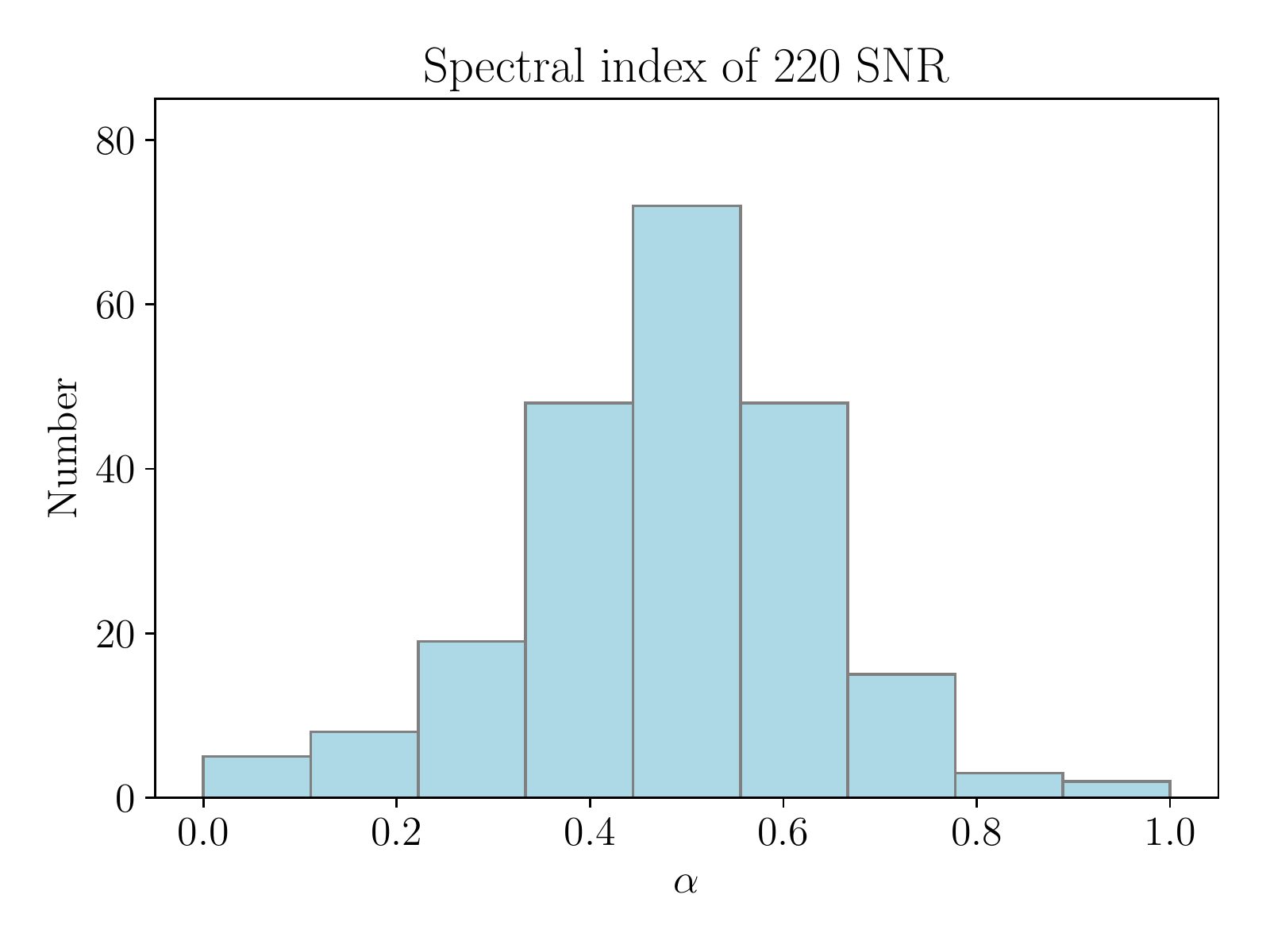}
  \caption{The distribution of the spectral index $\alpha$ of 220 SNRs in the Green catalog.}
  \label{fig:spectral_index_green}
\end{figure}

For the frequency range we are interested in, the photon spectrum of a SNR follows a simple power law:
\begin{align}\label{eq:snu_alpha}
  S_\nu  \propto E_\gamma^{-\alpha},
\end{align}
where $\alpha$ is the spectral index and $E_\gamma$ is the photon energy. We further choose the reference (or ``pivot'') frequency to be $1\;\mathrm{GHz}$, following the convention in the literature. The spectral luminosity is then given by:
\begin{align}
\label{eq:spectral_index}
  L_\nu =  S_\nu \; (4\pi D^2)
  = L_{1 \GHz} \bl( \frac{\nu}{1~\GHz} \br)^{-\alpha} , 
\end{align}
where $D$ is the distance of the source, and spherical symmetry of the source is assumed. From the Green catalog of SNRs~\cite{Green:2005yt,Green:2014cea,Green:2015isa,Green:2019mta}, one could see that spectral indices are centered around $\alpha \approx 0.5$ with considerable scattering, as shown in Fig.~\ref{fig:spectral_index_green}. We note that among the SNRs in the catalog, most are shell type, with less than 10 filled type SNR and about 20 of the composite type. The latter two types typically have a spectral index smaller than 0.5 but only constitute a small fraction of all the observed SNRs.

\subsection{Time evolution}
\label{sec:evolution}

An important input for our analysis is the time evolution of the spectral luminosity $L_\nu$ of SNRs in the radio band, also called radio light curves. While observed SNR light curves exhibit a variety of forms, they typically show an initial exponential rise and subsequent power-law decays. SNR light curves have been the subject of intense study for decades~\cite{Woltjer:1972,Dubner:2015aqa,Bietenholz:2020yvw}. Four distinct phases, first proposed in~\Refe{Woltjer:1972}, have been identified:
\begin{enumerate}
\item \textit{Free expansion phase:}
  in this phase, the swept-up gas mass is much smaller than the supernova mass. Sometimes it is also called the \textit{ejecta-dominated phase} \cite{1999ApJS..120..299T}. During this phase, the shock wave expands with a roughly constant velocity $\sim \mathcal{O}(0.01)$ (in units of the speed of light), leading to $r\propto t$, with $r$ the SNR radius, and $t$ the time elapsed after the explosion~\cite{2013tra..book.....W}. The optical depth due to external free-free absorption and synchrotron radiation drops over time, which results in an exponential rise in the luminosity shortly after the explosion \cite{1986ApJ...301..790W,Bietenholz:2020yvw,2013tra..book.....W}. This phase is estimated to last $\mathcal{O}(\mathrm{few}\times100)$ years by comparing the initial supernova mass and the mass swept up \cite{1986ApJ...301..790W,1999ApJS..120..299T,Draine2011jt,2013tra..book.....W}. 
  
\item \textit{Adiabatic expansion or Sedov-Taylor phase:} the energy loss rate is much smaller compared to the expansion rate, so energy conservation approximately holds. The driving force is the pressure behind the shock wave, with surrounding material being pushed away. Hydrodynamic estimates show an expansion as $r\propto t^{2/5}$ in this phase~\cite{2013tra..book.....W}. This phase can last up to $\sim \mathcal{O}(\mathrm{few} \times 10^4)$ years after the explosion \cite{Draine2011jt}. 
  
\item \textit{Snow plough or radiative phase:}
  after the remnant has dissipated half of its initial energy, the pressure driving the shock wave drops. The shell sweeps with a constant momentum and the expanding radius scales as $r\sim t^{1/4}$. This phase typically ends around $10^6$ years after the explosion \cite{Pavlovic:2012az,Draine2011jt}.
  
\item \textit{Dispersion phase:} the shock velocity falls below the speed of sound and the density drops to the same as that of the interstellar medium (ISM). It marks the merger of the SNR with the ISM. \end{enumerate}

Strictly speaking, one needs to implement full hydrodynamical simulations of the SNR time evolution for a precise computation of its properties. However, in our work, we only intend to provide a {\it proof-of-principle} computation of axion dark matter echoes triggered by SNRs. We will thus restrict ourselves to consider the contributions to the echo signal of the first two phases. In addition, we will apply some simple parametrizations of these two phases, which have been used in the literature.

In the first free expansion phase, we follow \Refe{Bietenholz:2020yvw} and take the following phenomenological model for the light curve:
\beq\label{eq:free}
    L_{\nu, \fr}(t) \equiv L_{\nu, \pk} ~ e^{\frac{3}{2}\left(1- {t_\pk}/{t}\right)} \bl( \frac{t}{t_\pk} \br)^{-1.5} \ ,
\eeq
where $L_{\nu, \pk}$ is the peak luminosity of the SNR, and $t_\pk$ is the corresponding peak time when $L_{\nu, \pk}$ is reached. Initially, the exponential rise dominates till the luminosity reaches its peak at $t_\pk$. Afterward, the light curve falls off following a power law. While there are large uncertainties associated with SNRs in general, typical value ranges for the peak parameters $L_{\nu, \pk}$ and $t_\pk$ can be derived from observations, as demonstrated in \Refe{Bietenholz:2020yvw}. After analyzing about 300 recent supernovae in the radio frequency $2 \text{\textendash} 10~\mathrm{GHz}$, the authors of \Refe{Bietenholz:2020yvw} found that the peak parameters are well described by a log-normal distribution, with means and standard deviations shown in \Tab{tab:pk}. However, these measurements cluster around $\sim 8~\mathrm{GHz}$, with a relatively narrow width (standard deviation of $\sim 2\ \mathrm{GHz}$). We therefore take the $L_{\nu, \rm pk}$ values quoted in \Tab{tab:pk} to be those for $L_{\rm 8GHz, \pk}$. Following \Eq{eq:spectral_index} we can deduce the peak luminosity at other frequencies, particularly 1 GHz, by applying the $\nu^{-\alpha}$ scaling.
\begin{table}[t]
   \centering
   \begin{tabular}{| c | c | c |}
      \hline
      Parameter & mean ($\mu$) & standard deviation ($\sigma$) \\
      \hline
      \hline
      $\log_{10} \bl( \frac{L_{\nu, \rm pk}}{\mathrm{erg \ s^{-1} \ Hz^{-1}}} \br)$ & $25.5$ & $1.5$ \\
      $\log_{10} \bl( \frac{t_{\rm pk}}{\mathrm{days}} \br)$ & $1.7$ & $0.9$ \\
      \hline
   \end{tabular}
   \caption{Mean and standard deviations of the SNRs peak light curve parameters $L_{\nu, \rm pk}$ and $t_{\rm pk}$ in the radio band, describing the free expansion phase; see \Eq{eq:free}. The parameters are log-normal distributed. While these parameters were derived from observations in a range of $2\text{\textendash}10~\mathrm{GHz}$, they are clustered around $8~\mathrm{GHz}$. Therefore, we take the $L_{\nu, \rm pk}$ values in this table to denote the spectral luminosity at $\nu = 8 ~ \mathrm{GHz}$; the values at other frequencies can be deduced following the power law $\nu^{-\alpha}$. For example, for a SNR with spectral index of $\alpha = 0.7$, its expected peak luminosity at 1 GHz (corresponding to the mean of the distribution) is $L_{\rm 1GHz, \pk} \approx 10^{26}~\mathrm{erg \ s^{-1} \ Hz^{-1}}$.}
   \label{tab:pk}
\end{table}

As the free expansion phase comes to an end, the SNR transitions into the adiabatic phase at a time $t_{\rm tran}$. The spectral luminosity follows a single power law during the adiabatic expansion as:
\bea\label{eq:adiab}
    L_{\nu, \ad}(t) & \equiv & L_{\nu, \tr} ~ \bl( \frac{t}{t_\tr} \br)^{-\gamma} \ , \nn\\
    L_{\nu, \tr} & \equiv & L_{\nu, \fr}(t_\tr) \ ,
\eea
with the adiabatic index \cite{2013tra..book.....W}
\beq\label{eq:gamma}
    \gamma  = \frac{4}{5} (2\alpha+1) > 0 \ .
\eeq

Alternatively, for SNRs still in the second phase, we can write the adiabatic evolution in terms of the SNR's spectral luminosity today, $L_{\nu, 0}$, as well as its age:
\bea\label{eq:adiab2}
    L_{\nu, \ad}(t) & \equiv & L_{\nu, 0} ~ \bl( \frac{t}{t_{\rm age}} \br)^{-\gamma} \ , \nn\\
    L_{\nu, \ad}(t_\tr) & = & L_{\nu, \fr}(t_\tr) \ .
\eea
Thus, the light curve of an SNR during its first two phases can be described entirely by anchoring the light curve with early luminosity 
$\{ L_{\nu, \pk}; t_{\pk}; t_{\tr}; t_{\rm age}; \gamma(\alpha) \}$, anchoring with late luminosity, $\{L_{\nu, 0}; t_{\pk}; t_\tr; t_{\rm age}; \gamma(\alpha) \}$, or any other combination such as luminosity at early and late times with the age being inferred, $\{ L_{\nu, \pk}; t_{\pk}; t_{\tr}; L_{\nu, 0}; \gamma(\alpha) \}$.

\begin{figure}[h]
  \centering
  \includegraphics[width=0.5\textwidth]{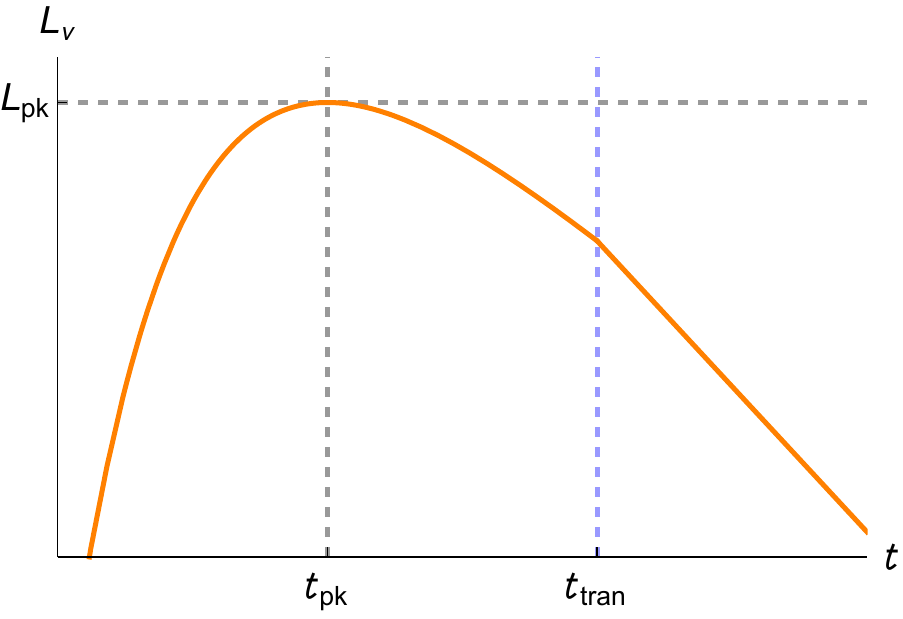}
  \caption{A schematic picture of the radio light curve of a SNR in our analysis, obeying \Eq{eq:master-eq-light-curve}. For $t \leq t_\tr$ the light curve follows \Eq{eq:free}; while for $t > t_\tr$ it follows \Eq{eq:adiab2} instead.}
  \label{fig:lightcurve}
\end{figure}

In summary, the model for the SNR light curve we will use throughout the rest of this paper is:
\bea
 L_\nu (t)
  & = &
    \begin{cases}
    L_{\nu, \fr}(t) \quad  t \leq t_\tr \\
    L_{\nu, \ad}(t) \quad \; \; t > t_\tr 
  \end{cases}
  \cr
  L_{\nu, \ad}(t_\tr)
  & = &
    L_{\nu, \fr}(t_\tr)  \ .
      \label{eq:master-eq-light-curve}
\eea
This is shown in Fig.~\ref{fig:lightcurve}. 
For the model to make sense, the light curve should reach the peak luminosity before its transition to the adiabatic phase, \ie $t_{\rm pk} > t_{\rm tran}$.  

\subsection{Supernovae rate}
\label{sec:sn-rate}

In our analysis, we will not only consider the SNRs that have already been observed and collected in the Green catalog but also make projections for possible SNRs that could be observed in the future. Thus, in this section, we will estimate the total number of SNRs in our galaxy. It is estimated that the supernovae formation rate is about 0.02 to 0.03 per year \cite{1994ApJS...92..487T}. Considering a time $\sim 10^5\;\mathrm{years}$, the time scale for a SNR going through the first three phases before merging with the ISM, there should be $2000 \sim 3000$ SNRs in the Milky Way. Yet there are only 294 SNRs that have been successfully identified so far~\cite{Green:2019mta}. Explanations of the deficit in observation and completeness of the SNR catalogs are discussed in \cite{Green:2005yt,Pavlovic:2012az,Green:2015isa,Long:2017seu}. 

We will further estimate the number of SNRs in our neighborhood. 
In Ref.~\cite{Green:2015isa}, the distribution of SNRs inside the Milky Way is estimated using an empirical two-dimensional density distribution:
\begin{equation}
  \Sigma(R)
  \propto
    \left (\frac{R}{R_\odot} \right )^a
    \mathrm{exp} \left ( -b \frac{(R-R_\odot)}{R_\odot} \right ),
    \label{eq:sigmaR}
\end{equation}
where $\Sigma(R)$ is the surface density of SNRs at radius $R$ from the galactic center and $R_\odot = 8.5 \;\mathrm{kpc}$ is the distance from the Sun to the galactic center. The best fit in Ref.~\cite{Green:2015isa} is
$a = 1.09, b = 3.87$. In Ref.~\cite{Case:1998qg}, however, the best fit is $a= 2.00 \pm 0.67, b = 3.53 \pm 0.77$ with the same fitting function.
In either case, for a total of 2000 SNRs with ages within $10^5$ years, we expect $\mathcal{O}(10)$ SNRs within 1~kpc of the Sun. 
As we will show that old SNRs nearby could potentially lead to strong signals, this estimate implies that we could look for echoes resulting from old SNRs as close as within $1~\mathrm{kpc}$.

\subsection{Extra-old SNRs}
\label{sec:extra-old-snrs}
In our analysis, we will consider SNRs as old as or even older than a few $\times 10^4$ years. We refer to these SNRs as extra-old SNRs. An extra-old SNR's evolution may no longer be in the free expansion or adiabatic phase. If an old SNR exits the adiabatic phase into the third phase, the light curve model in Eq.~\eqref{eq:master-eq-light-curve} could no longer apply to the subsequent evolution of its luminosity. However, these SNRs can still contribute significantly to the echo signals. This is because the signal is an integral of SNR's flux density when it was much younger.

There are two potential benefits from looking at extra-old SNRs. 
First, if an old SNR is further away from the galactic center than the Earth, their wavefronts pass the Earth and propagate toward a dark matter dense region, \ie close to the galactic center, which could enhance the integral in Eq.~(\ref{eq:snu_echo}).\footnote{On the other hand, the background increases close to the galactic center. One starts to lose sensitivity at frequency $\nu \lesssim 300 \;\mathrm{GHz}$ due to the optical depth $\tau \sim \mathcal{O}(1)$ if the echo comes precisely from the galactic center.}
Second, the extra-old SNRs have higher statistics than those still in the first two phases. Assuming a uniform formation rate of supernovae in a fixed volume around the Earth, we can expect 10 times more extra-old SNRs with age $\lesssim 10^5$ years than younger SNRs $<10^4$ years. 

Observationally, since an extra-old SNR is long past its peak time, there could be no historical record of its explosion. Direct observation of the radio emission from such a source today might be challenging. We may not have a complete catalog of all the old SNRs today. Therefore, we should look for echoes even if there is no observed SNR in the opposite direction. A positive identification of the echo signal from extra-old SNRs can in turn motivate searches for these missing SNRs.

\section{Detecting the axion echoes}
\label{sec:detection}

In this section, we will discuss the procedure to estimate the signal-to-noise ratio for the echo signal at a telescope on Earth. Detailed computations, which we will show later, demonstrate that the best sensitivity could be achieved in the radio band, due to a combination of large Bose enhancement to boost the signal and relatively low background. Thus we will focus on the radio telescopes. In the near future, one of the most powerful radio telescopes will be Square Kilometer Array (SKA). We choose SKA as a benchmark experiment to evaluate the prospect of detecting an echo signal triggered by SNRs.

SKA, with two sites in Australia and Africa, could provide over a million square meters of collecting area through thousands of connected telescopes when both phases (SKA1 and SKA2) are completed~\cite{SKA1summary}. Since SKA1 is being designed now, we will only focus on it for the rest of our work. SKA1 is further divided into two frequency ranges: SKA1-low and SKA1-mid. Currently, the setup of SKA1-low is designed to have~\cite{SKA1summary, SKA1report}:
\begin{itemize}
\item frequency range: 50 MHz -- 350 MHz;
\item number of stations: 512;
\item effective diameter of each station: 38 m;
\item longest baseline (separation between two stations): 80 km, 
\end{itemize}
while SKA1-mid has~\cite{SKA1report}
\begin{itemize}
\item frequency range: 350 MHz -- 15.4 GHz;
\item 133 dishes with a diameter of 15 m (SKA dishes), 64 dishes with a diameter of 13.5 m (MeerKAT dishes);
\item longest baseline: 150 km. 
\end{itemize}
For SKA1-low, each station will include 256 antennas, and work as a single telescope. For SKA1-mid, both types of dishes are offset Gregorian reflectors with four frequency bands. 

SKA could operate in both the single dish and the interferometer modes, each with its own advantages and disadvantages. We will consider both modes in our estimates and review their basic formulas below. 

\subsection{Single dish mode}
\label{sec:single-dish-mode}

For the single dish mode, the angle corresponding to the Rayleigh criterion, or equivalently the angular resolution of the telescope, is given by~\cite{2013tra..book.....W}
\begin{align}
  \label{eq:theta_res}
  \theta_{\rm res}
  =1.22 \left ( \frac{\lambda}{d} \right )  \, {\rm rad}
  \approx 1.4^\circ \left ( \frac{\mathrm{GHz}}{\nu}  \right )
  \left ( \frac{15~\mathrm{m}}{d} \right ) \, ,
\end{align}
where $\lambda$ is the wavelength and $d$ is the diameter of the dish. The angular size of the signal, $\theta_\ec$ is:
\beq\label{eq:theta_ec}
    \theta_\ec = {\rm max}(\theta_{\rm s}, \theta_{\rm ab}, 2\delta) \, ,
\eeq
where $\theta_{\rm s}$ is the angular size of the SNR source, $\theta_{\rm ab}$ is the aberration angle in Eq.~\eqref{eq:ab}, and $2\delta$, given by Eq.~\eqref{eq:extendedangle}, is the extended angular size to overcome the loss due to the peculiar motion of dark matter.

In the case when $\theta_\ec < \theta_{\rm res}$ (the inner structure of the signal will not be resolved in this case), the noise power of a single telescope is given by:
\begin{align}
  \label{eq:PnoiUnit}
  P_{\rm noi, \, unit}
  & =
    \frac{2\; T_{\rm sys} \Delta \nu}{\sqrt{2 \Delta \nu \, t_{\rm obs}}} = \sqrt{2} \; T_{\rm sys} \bl( \frac{\Delta \nu}{t_{\rm obs}} \br)^{1/2} \, .
\end{align}
Here the optimal search bandwidth $\Delta \nu$ is given in Eq.~\eqref{eq:bandwidth}; and $t_{\rm obs}$ is the observation time, for which we take 100 hours as a benchmark for both SKA1-low and SKA1-mid. The factor of 2 in the numerator accounts for the polarization degrees of freedom of the photon. The denominator accounts for the noise reduction due to independent measurements: $\Delta \nu \, t_{\rm obs}$ measurements in time, and those same 2 photon polarizations \cite{2013tra..book.....W,Caputo:2018vmy}. Note that this expression is independent of $\theta_\ec$. The total system noise temperature, $T_{\rm sys}$, has multiple contributions:
\beq\label{eq:Tsys}
    T_{\rm sys}  =  T_\cmb  + T_{\rm gal} + T_\atm + T_{\rm rcv} + T_{\rm spl} \ ,
\eeq
where $T_\cmb$ is the CMB temperature; $T_{\rm gal}$ is the synchrotron background from the Milky Way or extragalactic sources, evaluated {\it at the position of the echo in the sky}; $T_\atm$ is the atmospheric temperature; $T_{\rm rcv}$ is the receiver temperature; and $T_{\rm spl}$ is the spill-over temperature.\footnote{For the frequency range of SKA1, the atmospheric zenith opacity is small and thus we ignore the correction due to atmospheric absorption.} These temperatures could all depend on the frequencies being probed. We take $T_\cmb = 2.725$ K and ignore its small frequency dependence. $T_{\rm gal}$ is computed from the 408-MHz Haslam all-sky map~\cite{1981A&A...100..209H} by converting from 408~MHz to the frequency being observed.\footnote{We use the filtered Haslam map, where baseline striping and known strong point sources are removed from the map while faint point sources could still remain. Thus the faint point sources are taken into account in our background estimate.} 
This is done by using the power law fit, $T_{\rm gal} (\nu)= T_{408} (\nu/0.408)^p$, with exponent $p = -2.55$~\cite{1981A&A...100..209H} and frequency $\nu$ in unit of GHz.
Note that $T_{\rm gal}$ is the dominant contribution to $T_{\rm sys}$ for SKA1-low since it probes a lower frequency range. We infer $T_\atm$ from subtracting $T_\cmb$ and $\bar{T}_{\rm gal}$ ($T_{\rm gal}$ averaged over the sky probed by the telescope) from the total sky temperature (a sum of $T_\cmb$, $\bar{T}_{\rm gal}$ and $T_\atm$) provided in Fig. 19 of Ref.~\cite{SKA1report} for SKA1-low and in Fig. 6 of Ref.~\cite{2019arXiv191212699B} for SKA1-mid. For SKA1-low, $T_{\rm rcv} = 40$ K~(Fig. 9 of \cite{2015ExA....39..567D}). For SKA1-mid, the function of $T_{\rm rcv}$ for SKA dishes and MeerKAT dishes is provided in~\cite{2019arXiv191212699B}. For the overlapping frequency range of both types of dishes, we take a weighted mean of their $T_{\rm rcv}$ using their number of dishes as weights; we use instead $T_{\rm rcv}$ of the SKA dish for the rest of the frequency range. $T_{\rm spl}$ is set to be zero for SKA1-low and 3~K for SKA1-mid.

In the case when $\theta_\ec > \theta_{\rm res}$, the noise power of a single telescope is modified to be:\footnote{This is because when $\theta_{\rm echo} > \theta_{\rm res}$, it requires $n=(\theta_{\rm echo}/\theta_{\rm res})^2$ independent measurements to fully cover the signal. The noise rescales as $P_{\rm noi, unit} \propto \sqrt n$.}
\begin{align}
  \label{eq:PnoiUnit2}
  P_{\rm noi,\,unit}
  & =
    \sqrt{2}\; \bar T_{\rm sys} \left ( \frac{\Delta \nu}{ t_{\rm obs}} \right )^{1/2}
    \left ( \frac{\theta_{\ec}}{\theta_{\rm res}} \right ) \, ,
\end{align}
where $\bar T_{\rm sys}$ is the average system temperature over the solid angle of the signal $\propto \theta_\ec^2$. The enhanced signal collecting area expanded by the angle $\theta_\ec$ instead of $\theta_{\rm res}$ has two competing effects: it leads to an increase in both the noise power and the number of independent measurements by a factor of $\left ( \theta_{\ec}/\theta_{\rm res} \right)^2$. This results in the additional factor of  $\left ( \theta_{\ec}/\theta_{\rm res} \right)$ in the noise power in Eq.~\eqref{eq:PnoiUnit2}, compared to Eq.~\eqref{eq:PnoiUnit}. 
The total noise for $N$ telescopes operating in single dish (SD) mode is
\beq\label{eq:pnoiSD}
    P_{\rm noi;\, SD} = \sqrt{N} P_{\rm noi,\, unit} \ .
\eeq

On the other hand, the signal power is given by:
\begin{align}
  \label{eq:Psig}
  P_{\rm sig} = f_\Delta \, S_{\rm e} \; \eta \; A_{\rm tot} = \overline{S}_{\rm \nu_a, e} \, \Delta \nu \; \eta \; A_{\rm tot} \ ,
\end{align}
where $S_{\rm e}$ is the total axion echo flux (\ie irradiance), $\Delta \nu$ is the frequency bandwidth, $f_\Delta \approx 0.84$ the fraction of the flux that falls within that band (see \Eq{eq:bandwidth}, and \Refe{Ghosh:2020hgd}), $\overline{S}_{\rm \nu_a, e}$ the average flux \textit{density} (\ie \text{spectral} irradiance) over the bandwidth (see \Eq{eq:snu_echo}), $A_{\rm tot}$ the total geometric collecting area of the telescopes, and $\eta$ is the $\nu$-dependent detector efficiency. Note that the number of photon polarizations is already included in $S_{\rm e}$ and $\overline{S}_{\rm \nu_a, e}$, because of the phase space integral over the momentum measure $\DP$ (see \Sec{sec:axion-dark-matter}).

We extract $\eta$ from the sensitivities of SKA1 provided in \Refe{2019arXiv191212699B}.\footnote{More specifically, \Refe{2019arXiv191212699B} provides the sensitivity $\eta A_{\rm tot}/{\tilde{T}}_{\rm sys}$ as a function of frequency, where $\tilde{T}_{\rm sys}$ is the system temperature including the sky-averaged galactic temperature. Multiplying the sensitivity by ${\tilde{T}}_{\rm sys}/A_{\rm tot}$ gives us the efficiency $\eta$.} We show the efficiency in the left panel of \Fig{fig:baseline}. Finally, the signal-to-noise ratio in the single dish mode is simply given by:
\bea\label{eq:snSD}
    \left.{\rm s/n}\right |_{\rm SD} & = & \frac{P_{\rm sig}}{P_{\rm noi; \, SD}} \nn\\
    & = & \frac{\overline{S}_{\rm \nu_a, e}}{S_{\rm \nu, \, noi; \, SD}} \ ,
\eea
with $S_{\rm \nu, \, noi; \, SD}$ as the noise flux density for an array in single dish mode \cite{2017isra.book.....T}:
\beq\label{eq:SnuSD}
    S_{\rm \nu, \, noi; \, SD} \equiv \frac{ \sqrt{2} \; \overline{T}_{\rm sys}}{A_{\rm unit} \; \eta \; \sqrt{ N \;\Delta \nu \; t_{\rm obs}}} \; \max\bl( \frac{\theta_\ec}{\theta_{\rm res}}, 1 \br) \, ,
\eeq
with $A_{\rm unit} \equiv A_{\rm tot}/N$ as the average area of each individual telescope in the array.

\subsection{Interferometer mode}
\label{sec:interferometry-mode}

A single pair of telescopes (a two-element interferometer) has an angular resolution~\cite{2013tra..book.....W}  
\beq\label{eq:theta_b}
    \theta_b =\left( \frac{\lambda}{B}\right) \, {\rm rad} =  0.17^\circ \left ( \frac{\mathrm{GHz}}{\nu}  \right )
  \left ( \frac{100~\mathrm{m}}{B} \right ) \, ,
\eeq
where $B$ is the length of the interferometer baseline (the separation between the two telescopes). Since the baseline is usually much longer than the diameter size of a single dish, the interferometer usually has a much better angular resolution than the single dish. In addition, for $N$ telescopes, there are $N(N-1)/2$ pairs of two-element interferometers, which could improve the signal-to-noise ratio by an additional factor of $\sqrt{N}$, compared to the single dish mode. On the other hand, for an image with angular size $\theta_\ec > \theta_b$, the visibility function $R \propto \left[\sin(\pi \theta_\ec/\theta_b)\right]/(\pi \theta_\ec/\theta_b)$ falls off fast to almost zero~(Sec. 9.2.3.2 in Ref.~\cite{2013tra..book.....W}). Thus not all telescope pairs could contribute to the detection. In particular, those pairs of telescopes far separated from each other will not be sensitive to objects with large angular sizes. Therefore, for an extended echo signal, the single dish mode could still potentially perform better than the interferometer mode. 

\begin{figure}[t]
  \centering
  \includegraphics[width=.48\textwidth]{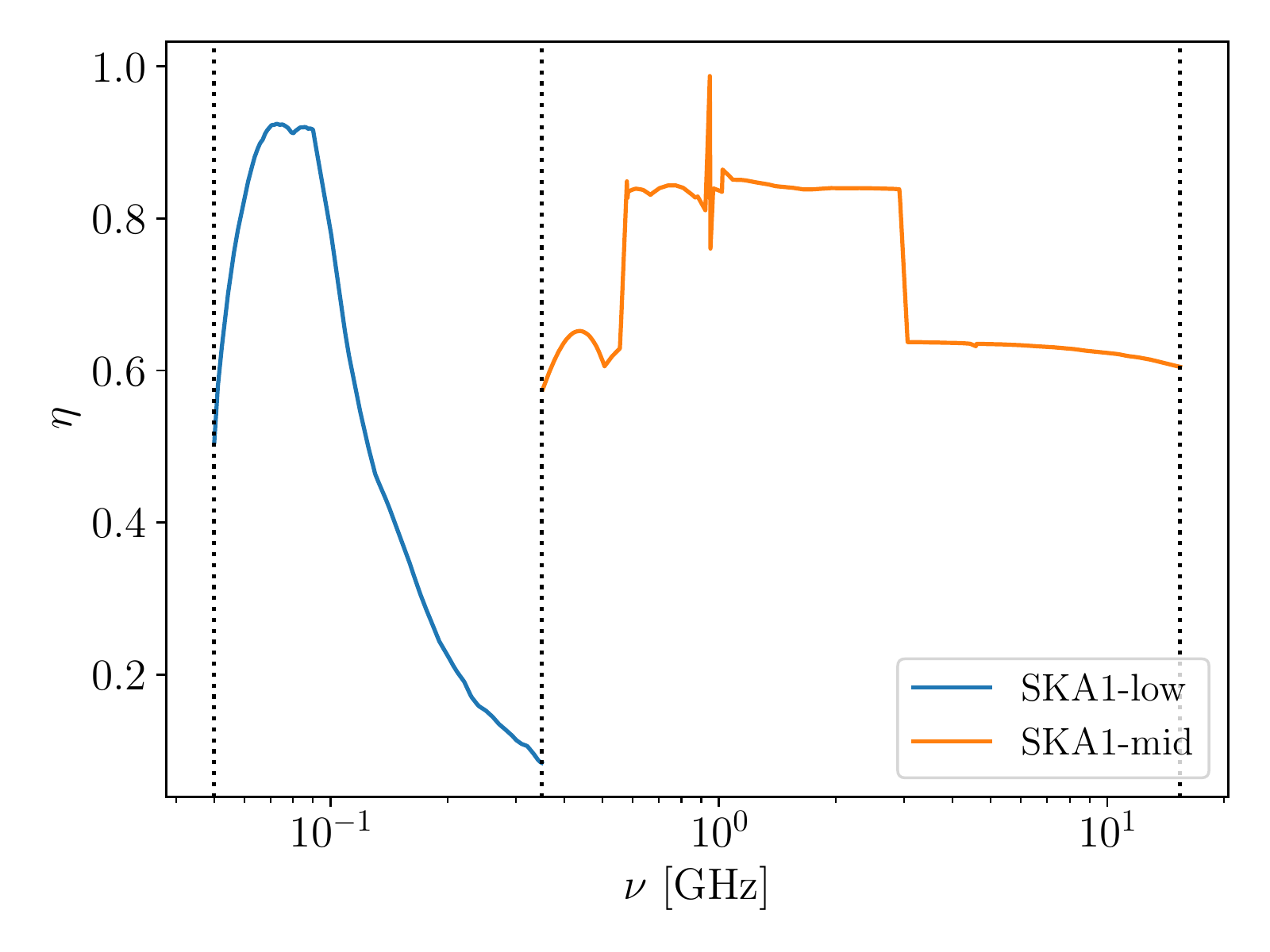}
  \includegraphics[width=.48\textwidth]{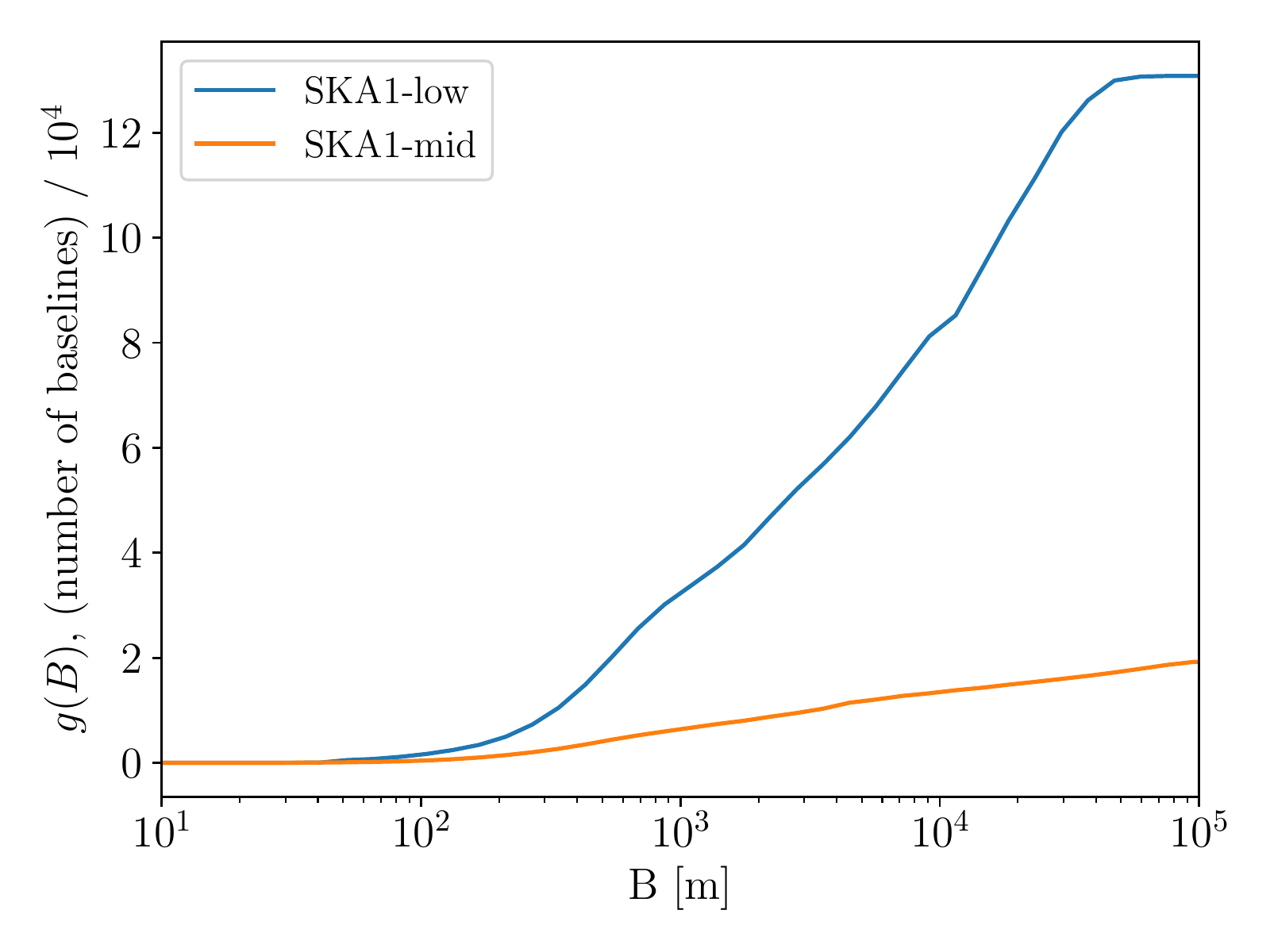}  
  \caption{\textit{Left}: efficiency of SKA1. The dotted lines from left to right correspond to the lowest frequency of SKA1-low we consider, frequency switching from SKA1-low to SKA1-mid, and highest frequency of SKA1-mid. \textit{Right}: baseline distributions of SKA1-mid and SKA1-low. The value of $g(B)$ represents the number of baselines shorter than $B$. }
  \label{fig:baseline}
\end{figure}

From the SKA1 design summary~\cite{SKA1summary}, we extract the configuration of the telescopes. We then count the baseline length of each pair to obtain a distribution of the baselines. We parametrize it with $g(B)$, the number of baselines shorter than $B$. This function is presented in the right panel of Fig.~\ref{fig:baseline}. As discussed above, a pair of telescopes contribute to the detection only if $B < \lambda/\theta_\ec$.
 
The RMS noise level for the interferometry (IN) mode is similar to \Eq{eq:PnoiUnit}, after correcting it by the increased noise from the multiple telescopes and dividing by the statistics of multiple independent measurements
\beq\label{eq:PnoiIN}
    P_{\rm noi; \, IN} = P_{\rm noi; \, unit} \frac{N}{\sqrt{N_{\rm pairs}}} \ ,
\eeq
where $N$ is the number of telescopes active in the interferometry mode and making up $N_{\rm pairs}$, the number of pairs of telescopes that contribute to the detection:
\beq\label{eq:Npairs}
N_{\rm pairs} = g( \lambda/\theta_{\rm echo}) \, .
\eeq
%
We denote the minimum baseline length in SKA1 $B_{\rm min}$ (if $B_{\rm min} >\lambda/\theta_\ec$, the interferometer mode will fail to work). The signal-to-noise ratio is:
\bea\label{eq:snIN}
    \left. {\rm s/n} \right|_{\rm IN} & = & \frac{P_{\rm sig}}{P_{\rm noi; \, IN}} \nn\\
    & = & \frac{\overline{S}_{\rm \nu_a, e}}{S_{\rm \nu, \, noi; \, IN}} \ .
\eea
where $S_{\rm \nu, \, noi; \, IN}$ is the noise flux density \cite{2017isra.book.....T}:
\beq\label{eq:SnuIN}
    S_{\rm \nu, \, noi; \, IN} \equiv \frac{ \sqrt{2} \; \overline{T}_{\rm sys}}{A_{\rm unit} \; \eta \; \sqrt{ N_{\rm pairs} \;\Delta \nu \; t_{\rm obs}}} \ .
\eeq
For a pointlike source, this implies a sensitivity equivalent to an antenna of aperture $\sqrt{N_{\rm pairs}} A \sim \sqrt{N(N-1)}A \sim N A$ when the number of telescopes $N$ is large.

\section{Results}
\label{sec:results}

Having already developed the formalism for the computation of the axion echo signal strength, as well as having established our treatment of the SKA telescope's properties and detection abilities, we devote this section to their application to a variety of SNRs. We will endeavor to demonstrate the effectiveness of this technique in axion dark matter detection efforts. To do this, we calculate the reach of the axion-photon coupling for various axion masses, focusing on SKA1 as an example radio telescope. We look first at the discovery potential of SNRs that are already observed, and afterward, we consider the reach projections of potential future SNRs.

\subsection{Constraints from the Green Catalog}
\label{sec:constr-from-greens}

We analyzed the SNRs in a catalog\footnote{\href{http://www.mrao.cam.ac.uk/surveys/snrs/}{\tt www.mrao.cam.ac.uk/surveys/snrs/}.} of 294 supernova remnants \cite{Green:2014cea,Green:2019mta} (henceforth the ``Green catalog'' or GC). We further supplemented the GC catalog with another SNR catalog\footnote{\href{http://www.physics.umanitoba.ca/snr/SNRcat}{\tt www.physics.umanitoba.ca/snr/SNRcat}.} \cite{2012AdSpR..49.1313F}, which includes age estimates for many SNRs also found in the GC. We then selected those SNRs from the GC with a known distance, flux density today, spectral index, and age, after which we are left with a total of 60 SNRs that can be used for our computation. For an SNR whose age is uncertain and known only within a range, we took the geometric mean of the upper and lower bounds as a benchmark value.

Subsequently, we calculated the SKA1 telescope signal-to-noise ratio of the SNRs' corresponding axion echoes in two separate cases:
\begin{enumerate}
 \item \textit{``Adiabatic-only'':} We considered contributions to the echo coming only from the portion of the line of sight corresponding to the adiabatic phase of the SNR's history, thereby excluding the free expansion phase. In this conservative case, the only free parameter is then the duration of the adiabatic phase or, alternatively, the age $t_{\rm tran}$ at which the SNR transitioned from the free to the adiabatic expansion, typically $\sim \mathcal{O}(100~\mathrm{years})$~\cite{2013tra..book.....W}.
 \item \textit{``Free + Adiabatic'':} We included contributions to the echo coming from the free expansion phase. The signal is then enhanced with respect to the previous case, at the cost of one more unknown parameter. Indeed, the echo signal depends on a choice of two out of three parameters $\{ t_{\rm tran}, t_{\rm pk}, L_{\rm \nu, \pk} \}$. The third one can be calculated in terms of the other two by using \Eq{eq:master-eq-light-curve}, for a SNR with known age and luminosity today. 
\end{enumerate}

\begin{table}[h]
   \centering
   \begin{tabular}{| c |}
      \hline
      G39.7-2.0 \\
      \hline
      \hline
      $(l, b) = (39.7\degree, -2\degree)$ \\
      $\theta_{\rm s} = 85$ arcmin \\
      $S_{\rm 1 GHz, s}^{(0)}= 85$ Jy \\
      $D = 4.9$ kpc ($4.5 \,\text{\textendash}\, 5.5$ kpc)\\
      $t_{\rm age} = 30,000 \,\text{\textendash}\, 100,000$ years \\
      $\alpha = 0.7$ ($0.5 \,\text{\textendash}\, 0.8$) \\
      $\gamma = 1.92$ ($1.6 \,\text{\textendash}\, 2.08$) \\
      \hline
   \end{tabular}
   \caption{Astrophysical properties of SNR G39.7-2.0. $S_{\rm 1 GHz, s}^{(0)}$ is the flux density of the source today, for which there is roughly $\sim 10\%$ uncertainty \cite{2011A&A...529A.159G,Green:2014cea,Green:2019mta}. In our computations we take the geometric mean of the age range given in \cite{2012AdSpR..49.1313F}, namely $\approx 5.5 \times 10^4$ years. For the distance $D$ and the spectral index $\alpha$, we use the values found in \Refe{Green:2014cea,Green:2019mta}, while for the adiabatic expansion index $\gamma$, we assume \Eq{eq:gamma}, the Sedov-Taylor formula. In parenthesis we quote the uncertainty ranges of these parameters, according to various studies \cite{Lockman:2007sz,Marshall:2013mka,Blundell:2008jd,Broderick:2018mcy}. For more details about the impact of these uncertainties on sensitivities, see \App{app}.}
   \label{tab:w50}
\end{table}

\begin{figure}[h]
  \centering
  \includegraphics[width=0.68\textwidth]{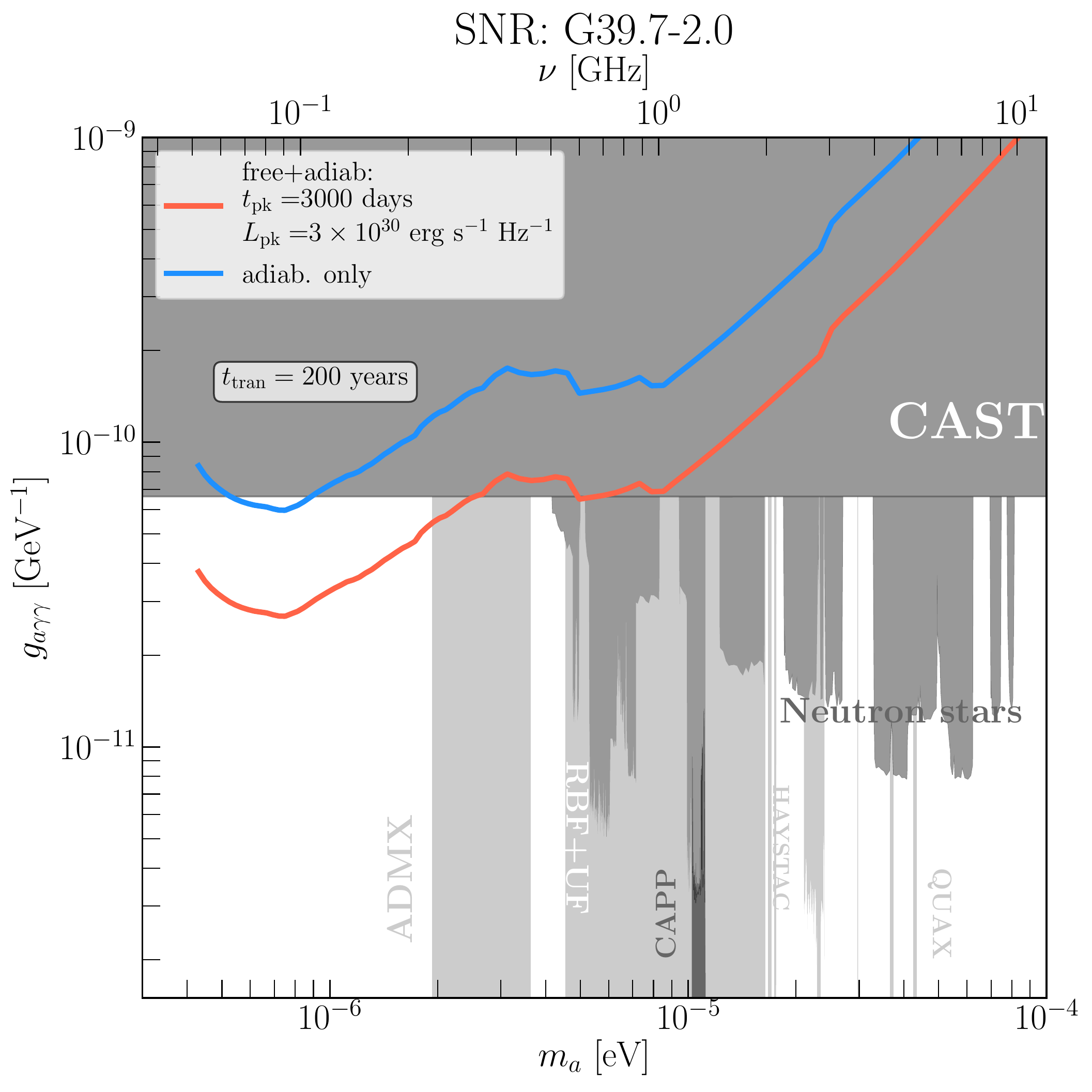}
  \caption{The sensitivity reach for axion dark matter coupling $g_{a\gamma\gamma}$ of SKA1, for the echo produced by SNR G39.7-2.0. We take a signal-to-noise ratio of $\mathrm{s/n}=1$. SKA1-low and SKA1-mid, in both their single dish and interferometer modalities, are combined into a single curve. We assume a typical value of $t_{\rm tran} = 200$ years. The reach for the conservative ``adiabatic-only'' case is shown in blue. The reach for the ``free+adiabatic'' case is plotted in red. For the latter case, we further assume $t_{\rm pk} = 3,000~\mathrm{days}$, which yields $L_{\rm 1GHz , \pk} = 3\times10^{30}~\mathrm{erg \ s^{-1} \ Hz^{-1}}$. The grey regions are existing bounds from previous literature: CAST \cite{CAST:2007jps,CAST:2017uph}, ADMX \cite{ADMX:2009iij,ADMX:2018gho,ADMX:2019uok,ADMX:2021nhd,ADMX:2018ogs,Bartram:2021ysp,Crisosto:2019fcj}, RBF+UF \cite{PhysRevLett.59.839,PhysRevD.42.1297}, CAPP \cite{Lee:2020cfj,Jeong:2020cwz,CAPP:2020utb}, HAYSTACK \cite{HAYSTAC:2020kwv}, QUAX \cite{Alesini:2019ajt,Alesini:2020vny}, and neutron stars \cite{Foster:2020pgt,Darling:2020uyo,Battye:2021yue}. These bounds are taken from \href{https://github.com/cajohare/AxionLimits}{\tt github.com/cajohare/AxionLimits}.}
  \label{fig:gc_reach}
\end{figure}

In \Fig{fig:gc_reach}, we show the expected SKA1 reach for the most promising SNR in the GC, G39.7-2.0 (also known as W50), whose properties are listed in \Tab{tab:w50}.
In this case, we anchor the light curve with its flux density today and extrapolate backward. We assume that the SNR is still in its adiabatic phase, therefore the light curve evolves according to Eq.~\eqref{eq:adiab2}. 
Since we are interested in demonstrating the sensitivity of the axion echo technique for axion dark matter detection by SKA, we show the reach setting the signal-to-noise ratio $\mathrm{s/n} = 1$ for illustrative purposes. In our figure, we have combined SKA1-low and SKA1-mid in a single curve, and have used either the single-dish or the interferometer mode, depending on which mode is the most sensitive at the particular frequency under consideration. We have assumed a typical value of $t_{\rm tran} \approx 200~\mathrm{years}$. The conservative case of adiabatic-only contribution to the axion echo is shown in blue, while the reach whose computation also included contributions from the free expansion is shown in red. For the latter, we have further assumed a peak time of $t_{\rm pk} = 3 \times 10^3~\mathrm{days}$, for which \Eq{eq:master-eq-light-curve} yields a peak luminosity of $L_{\rm 1GHz , pk} = 3\times10^{30}~\mathrm{erg \ s^{-1} \ Hz^{-1}}$.

From this figure we can see that SKA1 can be sensitive to echoes produced by axion dark matter decays stimulated by SNR G39.7-2.0 for a range of axion masses between $4\times 10^{-7}$ and $2\times 10^{-6}~\mathrm{eV}$, down to axion couplings of $g_{a\gamma\gamma} = 2 \times 10^{-11}~\mathrm{GeV^{-1}}$, a factor of $\sim 3$ below the current CAST limit. While the precise value of $g_{a\gamma\gamma}$ that a hypothetical discovery could give would depend on the SNR properties (in particular, during the early free expansion phase), the computation demonstrates the potential of the echo search strategy to probe axion dark matter. 

\begin{figure}[h]
  \centering
  \includegraphics[width=0.68\textwidth]{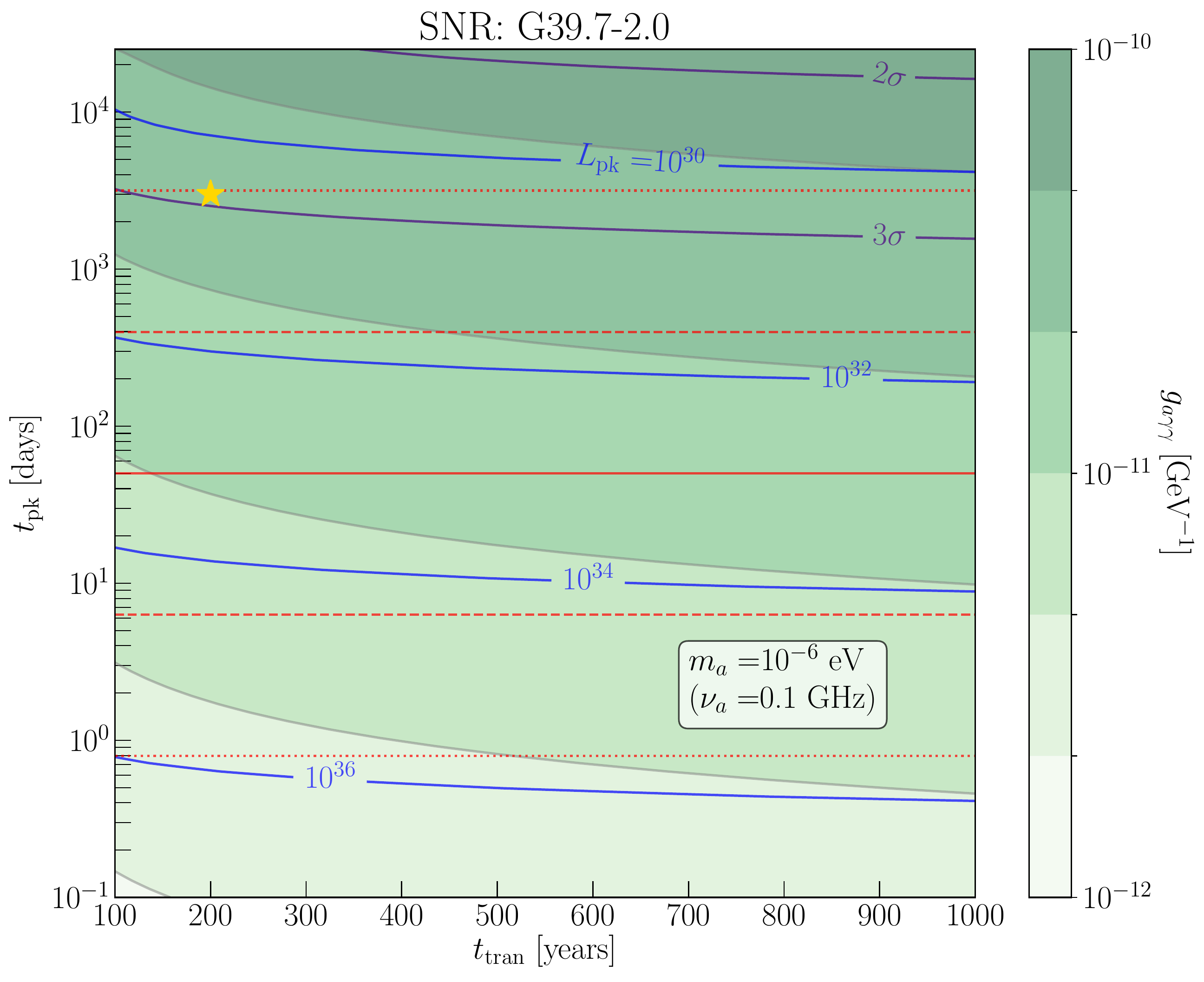}
  \caption{The sensitivity reach for axion dark matter coupling $g_{a\gamma\gamma}$ of SKA1 at $m_a = 10^{-6}~\mathrm{eV}$ ($\nu_a \approx 0.1~\mathrm{GHz}$), as a function of light curve parameters $t_{\rm tran}$ and $t_{\rm pk}$, for the echo produced by SNR G39.7-2.0. We take a signal-to-noise ratio of $\mathrm{s/n}=1$. The blue curves represent the values of $L_{\rm 1GHz , \pk}$ (in units of $\mathrm{erg \ s^{-1} \ Hz^{-1}}$) that satisfy the light curve equation \Eq{eq:master-eq-light-curve}. How typical these values are is represented by the purple curves, which denote the number of standard deviations from the mean of $L_{\rm 1GHz , \pk}$ \cite{Bietenholz:2020yvw}. The horizontal red lines in turn represent the mean (solid), $\pm 1 \sigma$ (dashed), and $\pm 2\sigma$ (dotted) values for $t_{\rm pk}$ (\textit{idem}). The golden star represents the light curve parameters assumed in \Fig{fig:gc_reach}.}
  \label{fig:ttr_tpk_reach}
\end{figure}

In \Fig{fig:ttr_tpk_reach}, we explore the effect that a different choice of values for $t_{\rm tran}$ and $t_{\rm pk}$ has on the SKA sensitivity to axion echoes, for an axion mass of $m_a = 10^{-6}\ \mathrm{eV}$ or, equivalently, a photon frequency of $\nu_a \approx 0.1 \ \mathrm{GHz}$. The green contours represent the $g_{a\gamma\gamma}$ sensitivity reach for $\mathrm{s/n} = 1$, and the golden star represents the choice of $t_{\rm tran}$ and $t_{\rm pk}$ used in \Fig{fig:gc_reach}. In blue we show the corresponding contours for the peak luminosity at 1 GHz, $L_{\rm 1GHz , pk}$. Their corresponding deviation from the typical SNRs values described in \Tab{tab:pk} are shown in purple, at $+2\sigma$ and $+3\sigma$ levels: it can be seen that G39.7-2.0 is an atypically bright SNR (around the peak time $t_{\rm pk}$ shortly after the explosion), which is why it is such a good candidate for axion echo discovery.\footnote{Note that the peak luminosities obtained in \Refe{Bietenholz:2020yvw} are for frequencies around $\nu \approx 8~\mathrm{GHz}$: we need to multiply our $L_{\rm 1GHz , \pk}$ by the conversion factor $\bl( \frac{\rm 8~GHz}{\rm 1~GHz} \br)^{-\alpha}$ in order to compute the corresponding $L_{\rm 8GHz , \pk}$ and obtain the standard deviations (see discussion in \Sec{sec:snr}).} {Note that G39.7-2.0 is not the brightest SNR today. In fact, its current flux density is only about $1/30$ of that of another SNR, Cassiopeia A, the brightest extrasolar radio source. It is the luminosity in the early times of G39.7-2.0 that makes it a more promising source than Cassiopeia A.}  The horizontal red lines in turn represent the mean (solid), $\pm 1 \sigma$ (dashed), and $\pm 2\sigma$ (dotted) values for $t_{\rm pk}$ (see \Tab{tab:pk}). It is evident from this figure that satisfying \Eq{eq:master-eq-light-curve} for typical values of $t_{\rm tran}$ and $t_{\rm pk}$ yields large $L_{\rm 1GHz , \pk}$ and even stronger sensitivities to $g_{a\gamma\gamma}$.

\subsection{Projection of unobserved SNR}
\label{sec:projections}

In this section, we want to first explore echo signals triggered by extra-old SNRs with ages above $10^4$ years, which are briefly discussed in Sec.~\ref{sec:extra-old-snrs}. They could be long past their peak times and have not been observed so far. But they could still potentially lead to observable axion echo signals, as we will show. In addition, we will also consider an opposite scenario: young supernova explosion that could happen close to us in the near future. In this case, if the peak flux density is very high (e.g., comparable or higher than that of a solar burst), it could also lead to an observable echo signal.

\begin{table}[t]
   \centering
   \begin{tabular}{|c|c|c|c|c|}
      \hline
     Benchmark & A (adiabatic only) & B (free+adiabatic)& C (old) & D (new) \\
      \hline
      \hline
     $(l, b)$ & $(64\degree, -0.1\degree)$ &$(64\degree, -0.1\degree)$  & $(175\degree, 5 \degree)$  & $(40 \degree, 0 \degree)$ \\
      $\theta_{\rm s} \;[{\rm arcmin}]$ & $48$ &$48$ & $16(*)$ & $1.0(*)$\\
     $S_{\rm 1 GHz , s}^{(0)}$ [Jy] &  $310$  &  $310$ & 6.3$(*)$ & $2.1\times 10^6(*)$ \\
     $L_{\rm 1 GHz , \pk}$ [cgs] & (n/a) & $2.5\times 10^{30}(*) $ & 1.2 $\times 10 ^{29}$ & 1.2 $\times 10 ^{29}$ \\
     $t_{\pk}$ [day] & (n/a) & $t_{\tr}/30$ & 50 & 50 \\
     $t_{\tr}$ [year] & 100 & 100 & 4.1 & 4.1 \\ 
      $D$ [kpc] & 1.9 & 1.9 & 0.5 & 0.5 \\
      $t_{\rm age}$ [year] & $35,000$  & $35,000$   &  $55,000$& $10$\\
      $\alpha$ &  $0.65$ &  $0.65$ & $0.65$ & $0.65$\\
      $\gamma$ & $1.84(*)$ & $1.84(*)$  &  $1.84(*)$ & $1.84(*)$ \\
      \hline
   \end{tabular}
   \caption{Astrophysical properties of three hypothetical SNRs/supernovae in the four benchmarks. Note that the SNR in benchmarks A and B are the same. For benchmark A, we only consider its adiabatic phase while for benchmark B, we consider both the free expansion and adiabatic phase. The quantities with an asterisk symbol means that they are derived based on other inputs. Note that $\theta_{\rm s}$ is the size of the SNR based on an estimate of its radius. Luminosity is given in the cgs unit of $\lumcgs$. 
   }
   \label{tab:benchmarks}
 \end{table}
 
 We consider the following four benchmark SNRs/supernovae shown in Table~\ref{tab:benchmarks}. In benchmark A and B, we take a SNR with similar properties to G6.4-0.1, an observed SNR in GC which could lead to even stronger echo signals, compared to G39.7-2.0 discussed in the previous section. But G6.4-0.1's signal is in the northern hemisphere and could not be observed by SKA1. The hypothetical SNR in the first two benchmarks are, however, taken to be located in the northern hemisphere so that the echo in the southern hemisphere can be observed by SKA1. We fix its current flux density to be the same as that of G6.4-0.1 and extrapolate backward, assuming the light curve model in Eq.~\eqref{eq:master-eq-light-curve}. In benchmark A, we only consider the signal from the adiabatic phase. In benchmark B, we include the free expansion phase, assuming $t_{\tr} = 30 \,t_{\pk}$.
 In benchmarks C and D, instead of fixing the flux density today and extrapolating backward, we model the first two phases of the expansion by anchoring the early stages of the light curves. We choose $L_{\rm 1GHz,\, \pk}$ and $t_{\pk}$ to be within the total $2\sigma$ range shown in \Refe{Bietenholz:2020yvw} and quoted in \Tab{tab:pk}. Note that in \cite{Bietenholz:2020yvw}, the analysis is implemented in the frequency range $4-10$~GHz except for SN1987A. As described in \Sec{sec:snr} and the previous subsection, we convert this to the fiducial frequency 1 GHz used here using the spectral index. The only differences in the inputs of benchmarks C and D are their ages and galactic coordinates.
In benchmark C, even though the echo comes from a region around the galactic center, it is still more than $5^\circ$ away from it. Therefore, we neglect the optical depth as discussed in Sec.~\ref{sec:extra-old-snrs}.
Note that the SNRs in benchmark A, B, and C are extra-old ones with ages above $10^4$ years while the SNR in benchmark D is a baby one, which is still in the very early stage of the free expansion phase shortly after the explosion.

\begin{figure}[th]
  \centering
  \includegraphics[width=.49\textwidth]{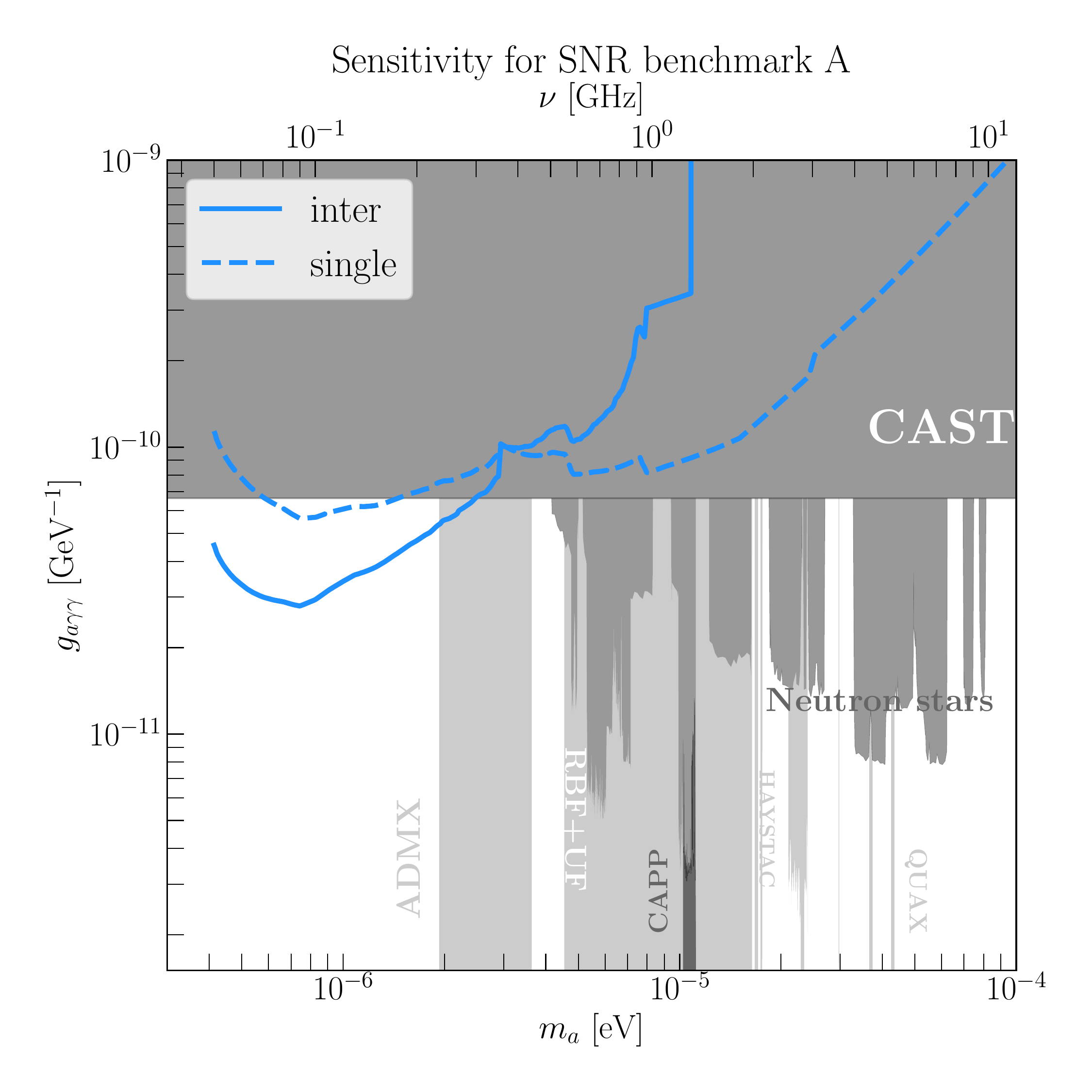}
  \includegraphics[width=.49\textwidth]{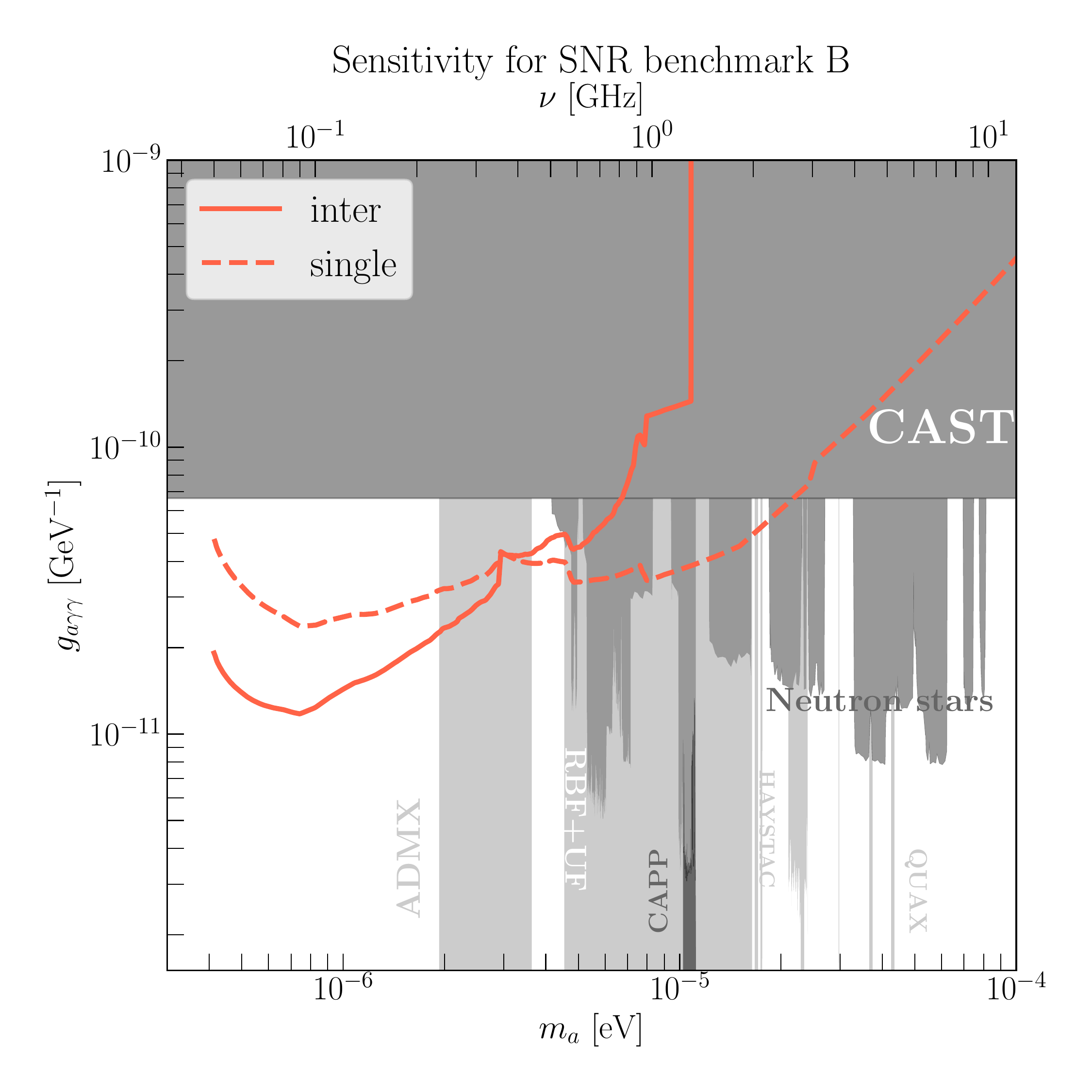}
  \includegraphics[width=.49\textwidth]{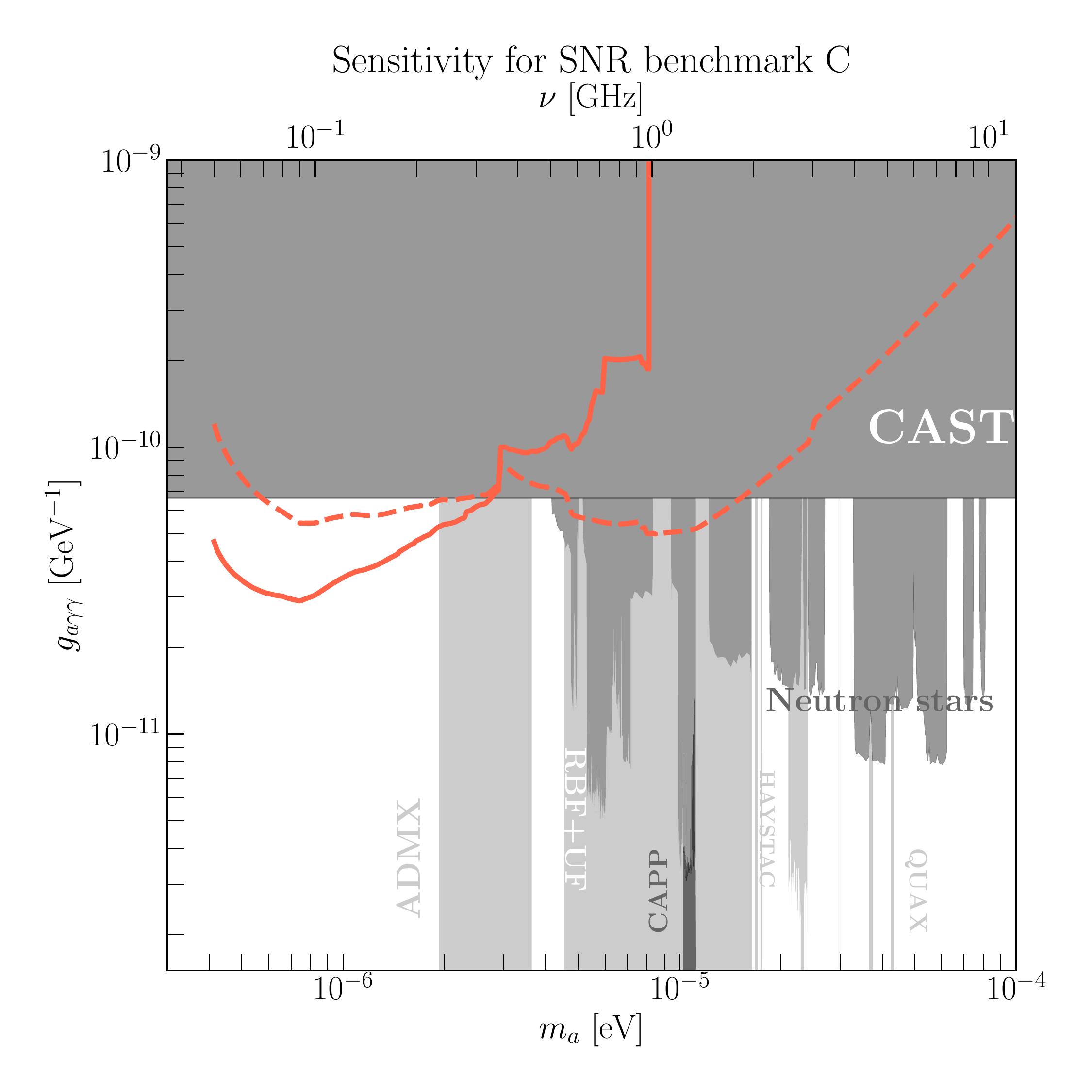}
  \includegraphics[width=.49\textwidth]{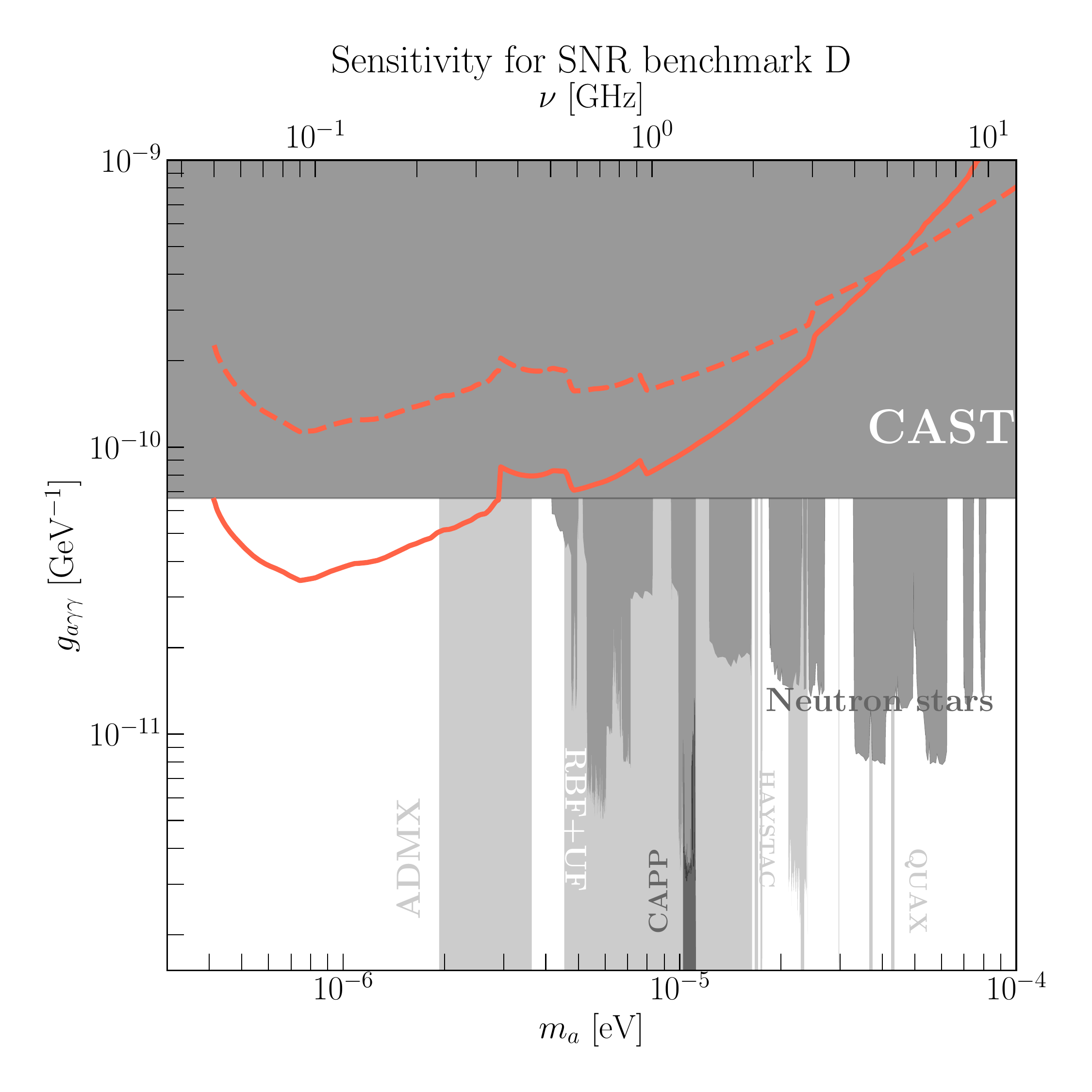}
  \caption{We show SKA1's sensitivities to axion DM for four benchmarks in Table~\ref{tab:benchmarks}. We set the signal-to-noise ratio, s/n$=1$. The blue curves only include the second phase (adiabatic expansion) of SNR, while the red curves also include the first phase (free expansion) as well. The solid (dashed) curves correspond to the interferometry mode (single-dish mode) of SKA1, respectively. In benchmarks A and B, the luminosity today is taken to be the same as SNR G6.4-0.1 in GC~\cite{Green:2014cea,Green:2019mta}; in benchmarks C and D, we choose $L_\pk$ and $t_\pk$ to be within the total 2$\sigma$ range based on recent supernova explosions \cite{Bietenholz:2020yvw}. See the main text for more details. The grey regions are the existing bounds: CAST \cite{CAST:2007jps,CAST:2017uph}, ADMX \cite{ADMX:2009iij,ADMX:2018gho,ADMX:2019uok,ADMX:2021nhd,ADMX:2018ogs,Bartram:2021ysp,Crisosto:2019fcj}, RBF+UF \cite{PhysRevLett.59.839,PhysRevD.42.1297}, CAPP \cite{Lee:2020cfj,Jeong:2020cwz,CAPP:2020utb}, HAYSTACK \cite{HAYSTAC:2020kwv}, QUAX \cite{Alesini:2019ajt,Alesini:2020vny}, and neutron stars \cite{Foster:2020pgt,Darling:2020uyo,Battye:2021yue}. These bounds are taken from \href{https://github.com/cajohare/AxionLimits}{\tt github.com/cajohare/AxionLimits}.}
  \label{fig:projection-benchmarks}
\end{figure}

For benchmarks C and D, we estimate the SNR/supernova radius as a function of their ages. During the free expansion phase, the speed of homologous expansion is estimated to be $v_{\rm hom}\approx 0.04 c$ \cite{1999ApJS..120..299T}. The radius is given by
\begin{align}
  R(t)
  & =
    \begin{cases}
      v_{\rm hom} t , & t < t_{\tr} \\
      v_{\rm hom} t_{\tr} \left (\dfrac{t}{t_{\tr}} \right )^{2/5} , & t \geq t_{\tr}  \, .
    \end{cases}
\end{align}

The sensitivities of these four benchmarks are presented in Fig.~\ref{fig:projection-benchmarks}. 
For all these benchmarks, we see that for the low frequency range (corresponding to $m_a$ around $10^{-6}$ eV), searching for echo signals at SKA1 could provide a better reach compared to CAST. A couple of comments are in order: 
\begin{itemize}
\item At low frequencies, the interferometry mode mostly dominates because of the lower array noise suppressed by $\sqrt{N_{\rm pairs}}$. At high frequencies, the resolution is enhanced, which leads to a suppressed visibility function as we discussed in Sec.~\ref{sec:detection}. 
  {As a result, the number of baselines that can observe the echo signal drops as the frequency increases. On the other hand, the number of dishes/stations that can observe the signal is constant in the entire frequency range of SKA1-low (SKA1-mid). One could verify that when the number of active baselines, $N_{\rm pairs}$, drops to be equal to that of dishes/stations, the two modes have equal sensitivities. At even higher frequencies, the single dish mode dominates. This explains the crossover of the two curves in all four plots in Fig.~\ref{fig:projection-benchmarks}.}
The single dish mode is almost always better at frequencies higher than 1~GHz except for baby supernovae as in benchmark D. For a very young supernova close by, which could be in its free expansion phase, it could have a small source size as well as small aberration angle and $\delta$ in Eq.~\eqref{eq:extendedangle} due to dark matter peculiar velocity. Thus the interferometry mode may not suffer from the degrading of the visibility function, which is present for more extended sources.
\item In benchmarks A and B, the peak luminosity is just above the $2\sigma$ range in \cite{Bietenholz:2020yvw}, just as SNR G39.7-2.0 discussed in the previous section. This may not be an issue since \Refe{Bietenholz:2020yvw} is a statistical study based on very recent supernovae explosions, while we consider a very old SNR in benchmarks A and B. As a result of selection bias, the samples of \cite{Bietenholz:2020yvw} may not be representative of the supernovae leading to old SNRs. On the other hand, in benchmark C, we choose the peak luminosity to be within the $2\sigma$ range in \Refe{Bietenholz:2020yvw}, and show that it could still lead to a strong echo signal, even though its luminosity today is at the tail of all well-measured SNRs \cite{Green:2019mta}. 
\end{itemize}

\begin{figure}[h]
  \centering
  \includegraphics[width=\textwidth]{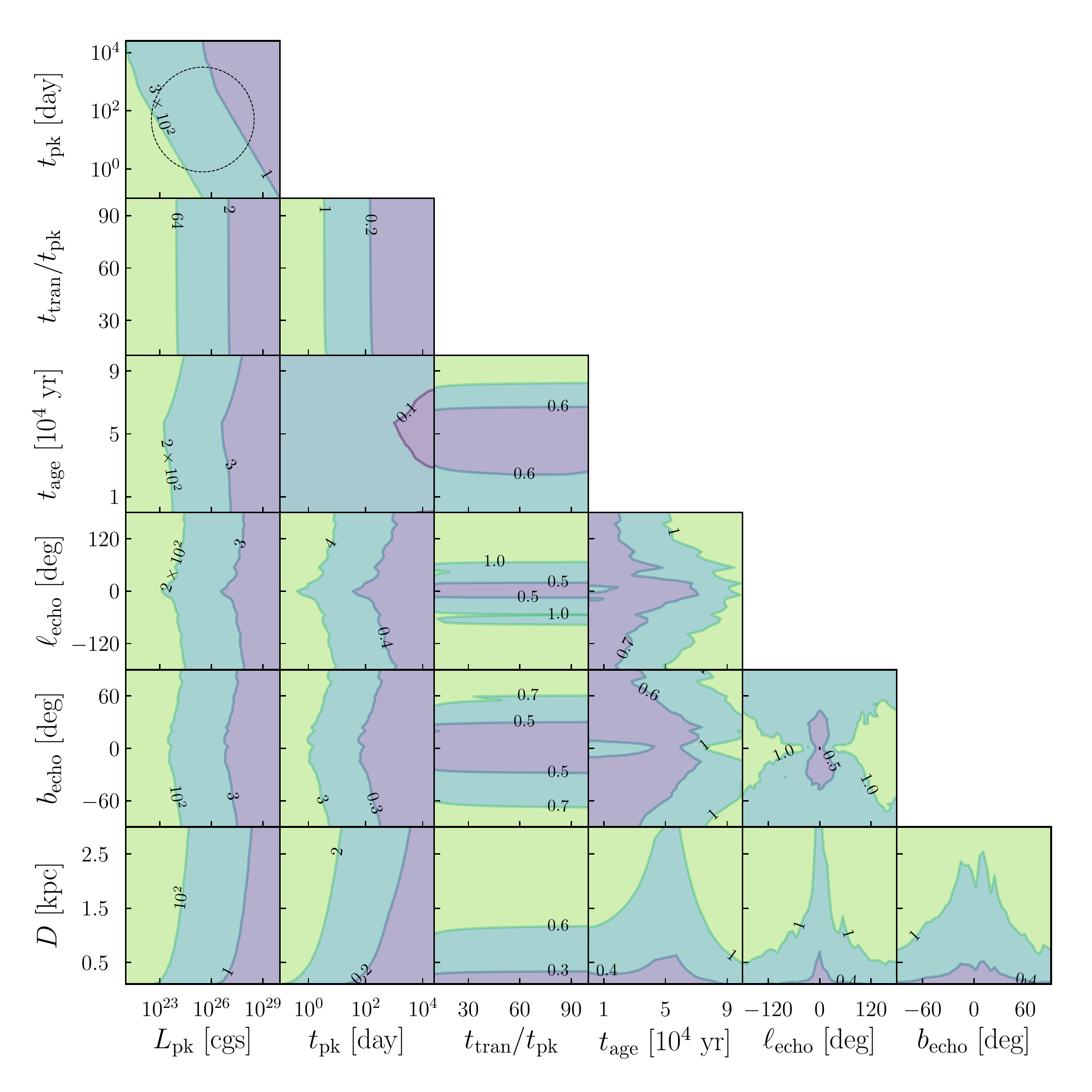}
  \caption{We show SKA1's reach of $g_{a\gamma\gamma}$ and its dependence on the SNR properties. We set the signal-to-noise ratio, s/n$=1$, and compute the sensitivity with fixed axion mass at $m_a = 10^{-6}\;\mathrm{eV}$ ($\nu \approx 0.1\;\mathrm{GHz}$). The contours are computed with both interferometry mode and single dish mode, with the former dominating at this frequency most of the time. The contours are labeled by $g_{a\gamma\gamma}/(10 ^{-10}\;\mathrm{GeV} ^{-1}$). 
The dashed circle in the top left corner block corresponds to the 2$\sigma$ contour of $(t_\pk, L_\pk)$ from \cite{Bietenholz:2020yvw}. Note that we vary two parameters in each panel, fixing all the other parameters to be the same as in benchmark C in Table~\ref{tab:benchmarks}. We consider the first two phases of the SNR to compute the signal strength.  The galactic coordinates are for the echo signal instead of the SNR. The small black dot at the center in the $\ell-b$ slice marks the galactic center, where the optical depth could modify the SKA1 reach. }
  \label{fig:slicing}
\end{figure}

Lastly, we show the dependence of the reach for $g_{a\gamma\gamma}$ on SNRs' intrinsic properties. We vary two parameters at a time and fix the other parameters to be the same as in benchmark C. We compute SKA1's reach of $g_{a\gamma\gamma}$ at $m_a = 10^{-6}\;\mathrm{eV}$. We show the $g_{a\gamma\gamma}$ contours in each two-dimensional slice of the parameter space. The results are collected in Fig.~\ref{fig:slicing}.
From the figure, we make the following observations:
\begin{itemize}
\item The signal is most sensitive to the peak luminosity, $L_\pk$, and the peak time, $t_\pk$.
\item An SNR nearby leads to a better reach, if the luminosity is fixed. This is expected as it is the flux density of the source (inversely proportional to the distance squared) that determines the signal flux density as shown in Eq.~\eqref{eq:snu_echo}. 
\item When the echo comes from the region around the galactic center, old SNRs lead to stronger signals. This is due to the wavefront of old SNR propagating into the dark matter-dense region of the galaxy. One should however ignore the reach at $\ell_{\rm echo} = b_{\rm echo}=0$, since the optical depth could be large from the galactic center.  
\item The reach has little dependence on the specific transition time, $t_{\rm tran}$, from the free-expansion phase to the adiabatic phase, when both phases are considered fixing $L_\pk$ and $t_\pk$. 
\end{itemize}

\section{Conclusions}
\label{sec:conclusion}

 In this paper we discuss the axion echo radio signal, namely the photons from axion stimulated decays that travel back to the radio source. We stress that there are two novel features of studying the echo signals compared to the forward-going radio signals also from axion stimulated decays: the low background, and its dependence on the history of the radio sources. The latter enables us to look for time-varying radio sources. We identify SNRs as good candidates and demonstrate that they could probe interesting parameter regions of axion dark matter. Even for the SNRs that are not the brightest radio sources we observe today, they can generate axion echoes stronger than the bright constant radio sources such as the radio galaxy Cygnus A. 

In discussing the candidate SNRs, we model their light curves with inputs from SNR catalogs as well as existing statistical studies based on newly exploding supernovae. We show the parameter region of $(m_a, g_{a\gamma\gamma})$ probed by SKA1 with known SNRs and with hypothetical but otherwise statistically normal ones that could be discovered in the future. We observe that SKA1 can reach $g_{a\gamma\gamma}$ as low as $\sim 10^{-11}\;\mathrm{GeV}^{-1}$ around $m_a \sim 10^{-6}\;\mathrm{eV}$. We also discuss the theoretical uncertainties in the modeling of the SNR light curves, the dependence of the SKA1 reach on SNR parameters, as well as the interplay with the two operation modes at SKA1.

There are several important directions to expand our work: 
\begin{itemize}
\item As we have emphasized, our work intends to provide a proof of concept estimate to demonstrate the power of searching for axion echo signals triggered by SNRs. More precise modeling of the SNR light curves could improve the estimate and put it on more solid ground. 
\item There is the very intriguing possibility that even for an SNR that has not been observed yet (\ie, it fails to pass the selection criteria of SNR searches), it could still lead to an observable echo signal. This could result in the interesting scenario of observing a radio line signal with no bright source in either direction along the line of sight as if the source is a ``ghost". In other words, searching for the echo signal could not only benefit the axion dark matter search but also the SNR search.  
\item We only focus on SKA1 as an example of a powerful radio telescope. One could extend the work to consider other radio telescopes. Note that with SKA1 to be constructed, studies of its potential in addressing fundamental questions and unveiling new physics will be useful and well timed. With SKA2's design currently under discussion, we leave it for future work to study its capability to detect axion dark matter in the mass range of $10^{-7}$~eV to $10^{-4}$~eV. 
\item 
  {The 408-MHz Haslam all-sky map still contains faint point sources that contributes to the background noise. With better measurements, the point sources could be removed in the future, which could potentially further suppress the noise and deepen the reach of SKA in $g_{a\gamma\gamma}$. }
\end{itemize}

{\textit{Note added --}
  During the late stages of the completion of this work, we became aware of other work~\cite{Sun:2021oqp} by Leung, Masui, Nambrath, Schutz, and Sun on the same subject. We note that the approaches and analyses of both works, while sharing common components, differ and complement each other in several aspects.}

\section*{Acknowledgements}

We thank Michael Bietenholz, 
{Kfir Blum,} Andrea Caputo, Michael Geller, Anson Hook, Gustavo Marques Tavares, Jonathan Pober, Matt Reece, Ben Safdi, Martin Schmaltz, Tomer Volansky, and Tien-tien Yu for useful discussions or correspondence. We also thank Yitian Sun, Katelin Schutz, Anjali Nambrath, Calvin Leung, and Kiyoshi Masui for coordinating with our research efforts, as well as for helpful discussions that have improved this work. M.B.A. would like to thank Stephanie Buen Abad for her unwavering support during the completion of this project. M.B.A. is supported by the NSF grant No.~PHY-1914731, and by the Maryland Center for Fundamental Physics (MCFP). J.F. is supported by the DOE grant No.~DE-SC-0010010 and the NASA grant No.~80NSSC18K1010. J.F. thanks the hospitality of Aspen Center for Physics (which is supported by National Science Foundation grant No.~PHY-1607611), where part of the work is implemented during workshop ``Dark Matter from the Laboratory to the Cosmos".  C.S. is supported by the Foreign Postdoctoral Fellowship Program of the Israel Academy of Sciences and Humanities, partly by the European Research Council (ERC) under the EU Horizon 2020 Programme (ERC-CoG-2015 - Proposal n.~682676 LDMThExp), and partly by Israel Science Foundation (Grant No.~1302/19). C.S. thanks the hospitality of INFN Galileo Galilei Institute for Theoretical Physics, where part of the work is finished during workshop ``New Physics from The Sky''.

\appendix
\section{Uncertainties in the properties of supernova remnant G39.7-2.0}\label{app}

In Sec.~\ref{sec:constr-from-greens}, we used the values of the distance and spectral index provided by the Green catalog \cite{Green:2014cea,Green:2019mta} for the most promising supernova, G39.7-2.0. For its age, we used the geometric mean of the upper and lower limits of the range of values given in \Refe{2012AdSpR..49.1313F}. All of these values are listed in \Tab{tab:w50}. Finally, we took a typical value of $t_\tr = 200$ years as the time at which the SNR transitioned from the free to the adiabatic expansion phase.

\begin{figure}[th!]
  \centering
  \includegraphics[width=.49\textwidth]{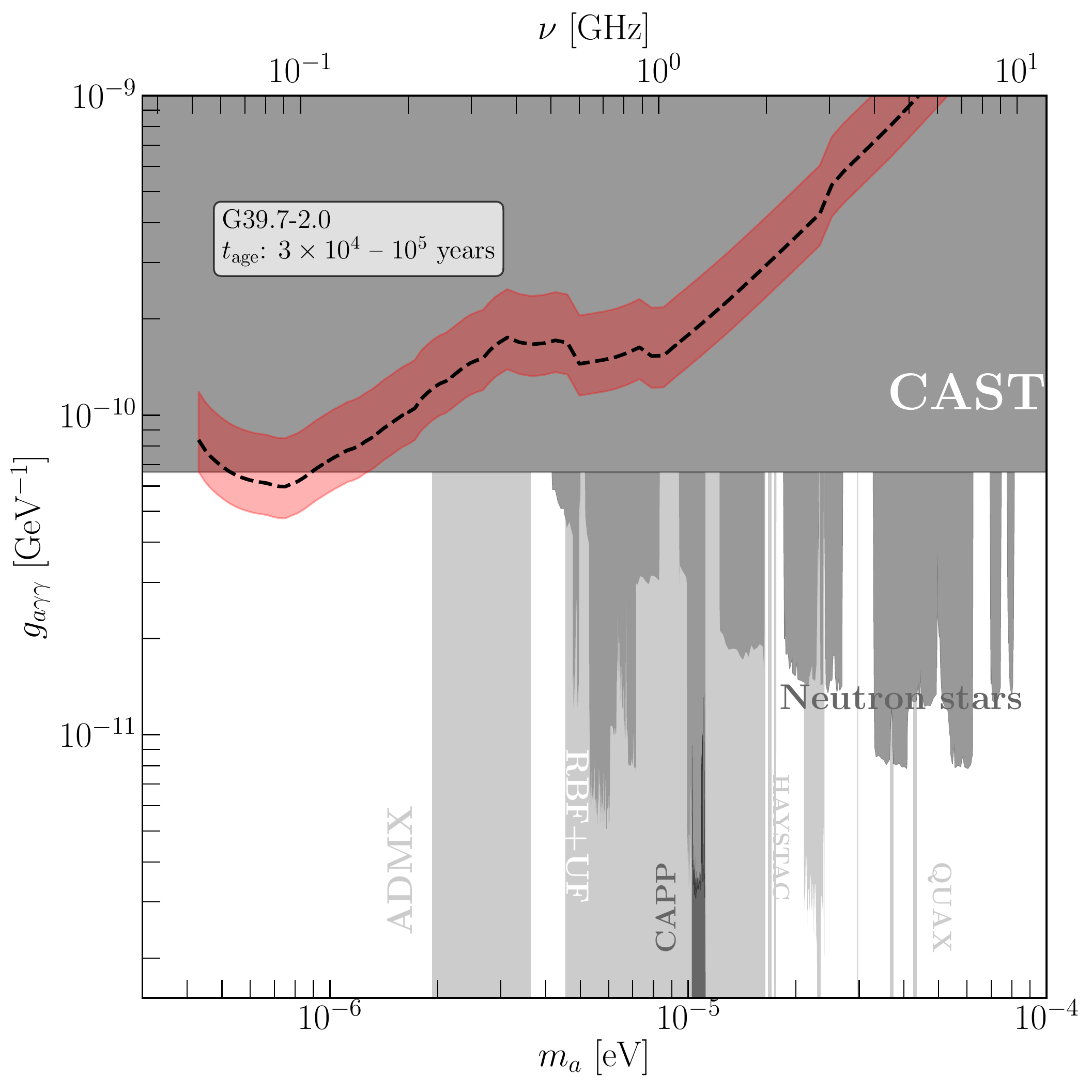}
  \includegraphics[width=.49\textwidth]{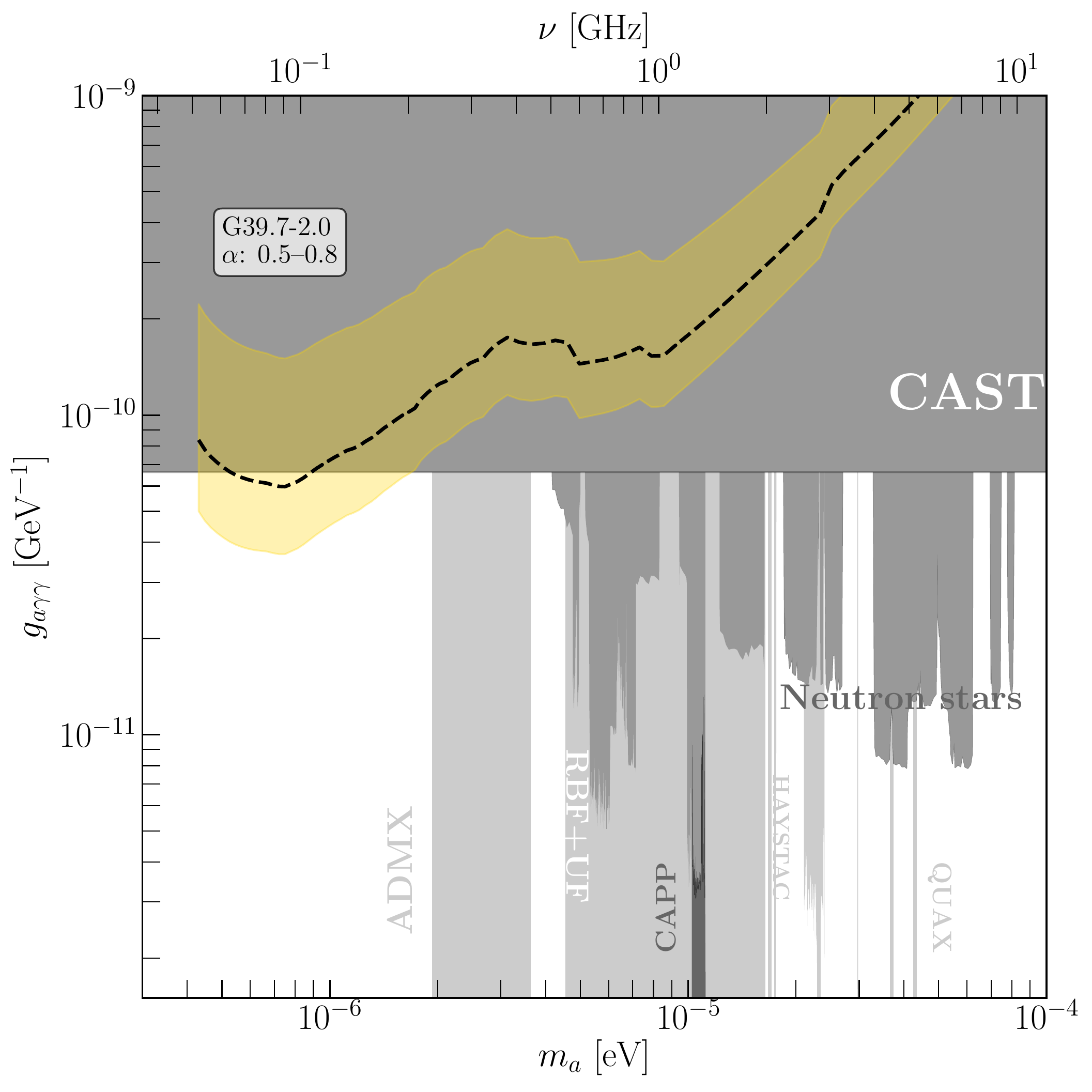}
  \includegraphics[width=.49\textwidth]{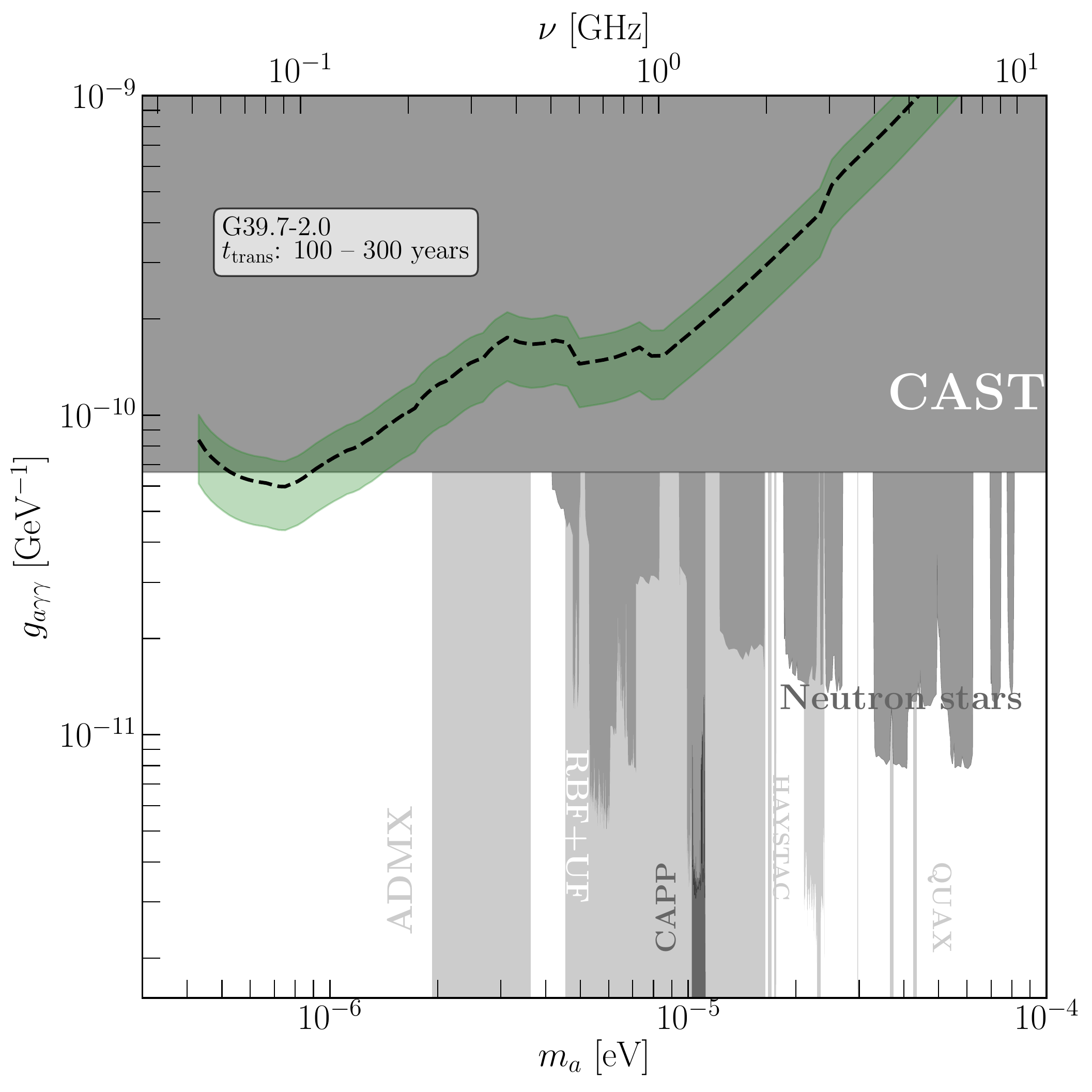}
  \includegraphics[width=.49\textwidth]{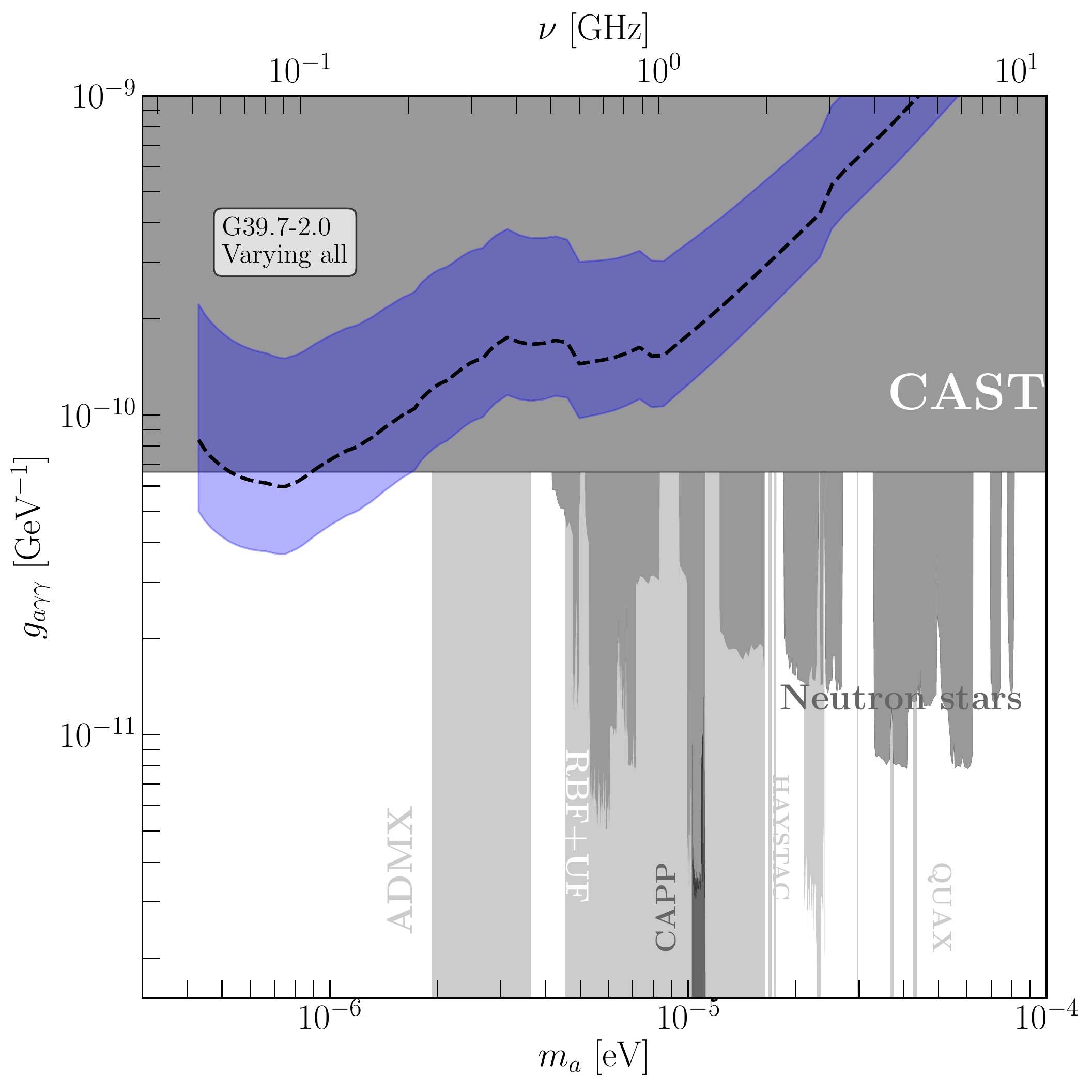}
  \caption{The uncertainties in the sensitivity reach for $g_{a\gamma\gamma}$ of SKA1, for the echo produced by SNR G39.7-2.0, as a result of the uncertainties of the SNR parameters. We take a signal-to-noise ratio of $\mathrm{s/n}=1$. SKA1-low and SKA1-mid, in both their single dish and interferometer modalities, are combined into a single curve. We take the conservative ``adiabatic-only'' case (\ie echoes produced by the adiabatic phase of the light curve only). In each panel, we vary a single parameter while keeping the other parameters fixed at the fiducial values used in the main text. \textit{Top left:} in red, varying $t_{\rm age}$ between $3\times10^4$ and $10^5$ years. \textit{Top right:} in yellow, varying $\alpha$ between $0.5$ and $0.8$. \textit{Bottom left:} in green, varying $t_\tr$ between $100$ and $300$ years. \textit{Bottom right:} in blue, the combined effect of the previous three cases; it is clearly dominated by the uncertainty in $\alpha$. The dashed black line in all four panels takes the same values we used in Sec.~\ref{sec:constr-from-greens}: $t_{\rm age} \approx 5.5\times10^4$ years, $\alpha = 0.7$, and $t_\tr = 200$ years. The grey regions are the existing bounds: CAST \cite{CAST:2007jps,CAST:2017uph}, ADMX \cite{ADMX:2009iij,ADMX:2018gho,ADMX:2019uok,ADMX:2021nhd,ADMX:2018ogs,Bartram:2021ysp,Crisosto:2019fcj}, RBF+UF \cite{PhysRevLett.59.839,PhysRevD.42.1297}, CAPP \cite{Lee:2020cfj,Jeong:2020cwz,CAPP:2020utb}, HAYSTACK \cite{HAYSTAC:2020kwv}, QUAX \cite{Alesini:2019ajt,Alesini:2020vny}, and neutron stars \cite{Foster:2020pgt,Darling:2020uyo,Battye:2021yue}. These bounds are taken from \href{https://github.com/cajohare/AxionLimits}{\tt github.com/cajohare/AxionLimits}.}
  \label{fig:reach-uncertain}
\end{figure}

In this appendix, we will describe the uncertainties of some parameters for G39.7-2.0, as well as their impacts on the sensitivity reach of SKA1. The reach curves in the axion parameter space resulting from varying these parameters independently are shown in \Fig{fig:reach-uncertain} as colored bands. Note that for  \Fig{fig:reach-uncertain}, we use only the adiabatic expansion phase of the light curve history in the computation of the echo signal (the conservative case we dubbed \textit{``adiabatic-only''} in the main body of this paper). As described in \Sec{sec:results}, the sensitivity could be improved with the inclusion of the echo flux from the free expansion phase of the light curve.

\begin{itemize}
 
\item \textbf{Flux density ($S_\nu^{(0)}$):} The value of $S_\nu^{(0)}$ we used in Sec.~\ref{sec:constr-from-greens} is taken from the Green catalog \cite{Green:2014cea,Green:2019mta}, and is $85~\Jy$ at $\nu = 1~\GHz$. Reference~\cite{2011A&A...529A.159G} quantifies the uncertainty in the flux density of this SNR and it is roughly $\sim10\%$ for measurements at wavelengths of $6$, $11$, and $21 \ \cm$ (frequencies of $5$, $2.7$, and $1.4\ \GHz$ respectively). Since the echo signal scales linearly with the flux density (\Eq{eq:snu_echo}) and, as we shall see below, the impact from uncertainties in other quantities far outweighs that from a $10\%$ uncertainty of $S_\nu^{(0)}$, we will not consider this effect any further.
 
 \item \textbf{Distance ($D$):} The Green catalog quotes $D=4.9~\kpc$ \cite{Green:2014cea,Green:2019mta}. Recent studies place the exact value between $4.5~\kpc$ and $5.5~\kpc$ \cite{Lockman:2007sz,Marshall:2013mka,Blundell:2008jd}. Since, however, it is the flux density $S_\nu$ that is relevant for the computation of the axion echo (\Eq{eq:snu_echo}) and of the signal strength (\Eqs{eq:snSD}{eq:snIN}), an uncertainty in $D$ merely translates into an uncertainty in the spectral luminosity $L_\nu$. We therefore do not consider the effect of this uncertainty any further.
 
 \item \textbf{Age ($t_{\rm age}$):} \Refe{2012AdSpR..49.1313F} and the public catalog SNRcat\footnote{\href{http://www.physics.umanitoba.ca/snr/SNRcat}{\tt www.physics.umanitoba.ca/snr/SNRcat}.} cite an age between $3\times10^4$ and $10^5$ years. This age roughly coincides with the typical end of the adiabatic expansion and the beginning of the snow plough phase (\Refe{Draine2011jt} cites a benchmark of $\sim 5 \times 10^{4}$ years). During the snow plough phase, the SNR rapidly loses energy, which translates into a weaker stimulation of axion decays and consequently a weaker signal. Nevertheless, since the estimated age is right around the transition between both phases, and in light of the larger uncertainty in the spectral index (see next item), we refrain from modeling the snow plough phase (which will complicate the SNR light curve model by introducing more parameters). In the top left panel of \Fig{fig:reach-uncertain}, we show in red the impact of the uncertainty in the age of SNR G39.7-2.0, considering echoes from the adiabatic phase only. The dashed black line corresponds to our fiducial age of $t_{\rm age} \approx 5.5 \times 10^4$ years, the geometric mean of the upper and lower ends of the age. 
  
 \item \textbf{Spectral index ($\alpha$):} The Green catalog quotes $\alpha = 0.7$ \cite{Green:2014cea,Green:2019mta}, but its precise value could lie between $0.5$ and $0.8$ \cite{Broderick:2018mcy}. Since $\alpha$ is an exponent, the uncertainty in its value has the largest impact on the sensitivity reaches. It not only affects the dependence of the reach on the axion mass (since $S_\nu \propto \nu^{-\alpha}$ as shown in \Eq{eq:snu_alpha}, and $\nu = m_a/4\pi$), but it also directly impacts the adiabatic expansion index $\gamma$, which according to the Sedov-Taylor formula (\Eq{eq:gamma}) is given by $\gamma = \frac{4}{5}(2\alpha + 1)$. The uncertainty in $\alpha$ then translates into an uncertainty in $\gamma$, which lies between $1.6$ and $2.08$. Since the axion echo is an integral over the SNR history (\Eq{eq:snu_echo}), the value of the adiabatic index is crucial in determining whether the echo signal is observable. The effect that the uncertainty in $\alpha$ has on the sensitivity reach is shown as the yellow band in the top right panel of \Fig{fig:reach-uncertain}; the dashed black line takes the Green catalog value of $\alpha = 0.7$ and corresponding $\gamma = 1.92$. 
 
 \item \textbf{Transition time ($t_\tr$):} The time at which the SNR transitioned from the free expanding phase to the adiabatic phase is typically of the order of $\mathcal{O}(\mathrm{few}\times 100)$ years (\Refe{Draine2011jt} quotes a benchmark of $\sim 190$ years). Without a dedicated study of the astrophysical properties of SNR G39.7-2.0 (possibly including astronomical observations and numerical modeling of the supernova remnant evolution), it is hard to know with precision at what time this transition roughly took place. In the simplest analytical estimates, $t_\tr$ is a function of the interstellar medium number density, the energy of the supernova explosion, and the mass of the ejecta \cite{1999ApJS..120..299T, Draine2011jt}. We can however quantify the impact of this uncertainty on the sensitivity reach by simply varying $t_\tr$ as a free parameter. In the bottom left panel of \Fig{fig:reach-uncertain}, we show in green the resulting reach band, allowing $t_\tr$ to vary between $100$ and $300$ years; the dashed black line assumes $t_\tr = 200$ years, as was used in \Fig{fig:gc_reach}. 
\end{itemize}

Finally, in the bottom right panel of \Fig{fig:reach-uncertain} we show in blue the sensitivity band resulting from the combination of all these variations. This band serves as a quantitative proxy of the uncertainty in the sensitivity reach for SNR G39.7-2.0. Clearly, the uncertainty in the spectral index $\alpha$ (and consequently on the adiabatic index $\gamma$) has the dominant effect. A more precise determination of the spectral index of G39.7-2.0 and of its expansion history could improve estimating the sensitivity of SKA1 to axion echoes.

\bibliography{axion_laser}
\bibliographystyle{utphys}

\end{document}